\documentclass[]{aastex631}
\usepackage{amsmath}
\usepackage[normalem]{ulem}
\usepackage{mathtools}

\newcommand{\modch}{\texttt{model\_CH}}
\newcommand{\modqs}{\texttt{model\_QS}}
\newcommand{\fexv}{{\ion{Fe}{15}}}
\newcommand{\feix}{{\ion{Fe}{9}}}
\newcommand{\ov}{{\ion{O}{5}}}
\newcommand{\oiv}{{\ion{O}{4}}}
\newcommand{\neviii}{{\ion{Ne}{8}}}
\newcommand{\siiv}{{\ion{Si}{4}}}
\newcommand{\bmag}{$\vert B_{z,phot}\vert$}

\newcommand{\review}[1]{#1}
\newcommand{\reviewtwo}[1]{#1}
\newcommand{\reviewthree}[1]{#1}
\newcommand{\reviewfour}[1]{{#1}}
\newcommand{\reviewfive}[1]{{#1}}
\newcommand{\reviewsix}[1]{{#1}}

\newcommand{\ppmg}{Paper~\rm{II}}
\newcommand{\ppsi}{Paper~\rm{I}}

\received{}
\revised{}
\accepted{}

\submitjournal{ApJ}
\begin{document}

\title{Comparison of plasma dynamics in Coronal Holes and Quiet Sun using flux emergence simulations}

\author[0000-0002-9253-6093]{Vishal Upendran}
\affiliation{SETI Institute, Mountain View, CA, USA - 94043}
\affiliation{Lockheed Martin Solar and Astrophysics Laboratory, Palo Alto, CA, USA - 94304}
\affiliation{Inter University Centre for Astronomy and Astrophysics, Pune, India - 411007}

\author[0000-0003-1689-6254]{Durgesh Tripathi}
\affiliation{Inter University Centre for Astronomy and Astrophysics, Pune, India - 411007}

\author[0000-0001-5424-0059]{Bhargav Vaidya}
\affiliation{Indian Institute of Technology - Indore, Simrol, India - 452020}

\author[0000-0003-2110-9753]{Mark C. M. Cheung}
\affiliation{CSIRO, Space \& Astronomy, Marsfield, NSW 2122, Australia}

\author[0000-0001-5457-4999]{Takaaki Yokoyama}
\affiliation{Astronomical Observatory, Kyoto University, Kyoto, Japan - 606-8502}

%--------------------
\begin{abstract}
This paper presents a comparison of~\review{plasma dynamics in} Coronal Holes (CHs) and Quiet Sun (QS) through 2.5D MHD flux emergence simulations. The magnetic reconnection between the emerging and the pre-existing flux leads to the formation of cool, dense plasmoids with hot boundaries, and hot \reviewtwo{\&} cool jets with velocities $\approx50$ km s$^{-1}$. We perform spectral synthesis in spectral lines~\review{probing transition region and coronal temperatures}. CHs show reduced intensities, excess upflows (downflows), and widths during the jetting (downflow) period~\review{when compared to QS}. During the ~\reviewtwo{jetting and downflow} periods, velocity and line width of the~\reviewtwo{hot} spectral lines in CHs show a strong positive correlation with the vertical magnetic field at z = 0, while the intensity of the cooler lines shows a weak correlation,~\review{which is not seen in QS}.  During the jetting period in CH, we find upflows in {\siiv} to be correlated (anti-correlated) with upflows (downflows) in other lines, and downflows in CH in {\siiv} to be correlated (anti-correlated) with upflows (downflows) in other lines~\review{when compared to QS}. During downflow, we find no strong correlation between {\siiv} and other line velocities. The correlation during the jetting period occurs due to coincident, co-spatial origins of the hot and cool jet, while the lack of correlation during the downflow phase suggests a decoupling of hot and cool plasma. These results demonstrate that flux emergence and reconnection with pre-existing flux in the atmosphere support a unified scenario for solar wind formation and coronal heating. 
\end{abstract}


%=======================
\section{Introduction} \label{sec:intro}
%=======================
The solar atmosphere depicts a wide variety of processes and morphological features. In the extreme ultraviolet (EUV) and X-ray observations of the solar corona, we observe dark structures called Coronal Holes (CHs) in contrast to the diffuse Quiet Sun (QS) emission. The CHs are well-known sources of solar wind, showing signatures of excess plasma upflows and spectral line widths when compared to QS \citep{Cranmer_2009}. These excess upflows are observed in the spectral lines forming at temperatures with $\log T/[K]\gtrapprox 5.7$ \citep{Peter_1999}, and are known to show association with the underlying network regions \citep{Hassler810}. The QS, on the other hand, shows excess emission at these temperatures when compared to CHs. However, using the observations recorded by Solar Ultraviolet Measurements of Emitted Radiation~\citep[SUMER;][]{wilhelm1995sumer}, \cite{Stucki_UVLinesSumer_Chs,Stucki_UVLinesSumer_correlations} have shown that in the photosphere and chromosphere, CHs and QS regions exhibit similar intensities, velocities, and line widths. Thus, CHs and QS, two regions similar low in the solar atmosphere, differentiate into regions with excess plasma upflows and excess local heating respectively. Hence, it becomes important to study the salient properties of these two regions as a function of height, in the context of solar wind emergence and coronal heating.

The CH and QS regions are known to exhibit statistical differences when the underlying photospheric magnetic flux density ($\vert B\vert$) is taken into consideration~\cite[][henceforth {\ppsi}]{TriNS_2021}. \review{To this end, {\ppsi} considers the intensity, velocity, and line widths of the {\siiv} 1394~{\AA} line recorded by the Interface Region Imaging Spectrometer~\citep[IRIS;][]{iris}, binned with the underlying $\vert B\vert$.} {\ppsi} demonstrates that the CHs exhibit lower intensities when compared to QS regions , when the intensities are binned in $\vert B\vert$.~\reviewtwo{These differences were found to increase with $\vert B\vert$}.They also find excess upflows (downflows) in CHs (QS) when the red- and blue-shifted pixels are taken separately, and binned in $\vert B\vert$. Furthermore, the non-thermal widths were found to remain almost the same in the two regions. \review{Importantly, the intensity, velocity and non-thermal widths were found to increase with increasing $\vert B\vert$, in {\ppsi}}.

These statistical differences between CH and QS have also been seen further lower in the atmosphere. \cite{PradeepKashyap2018} found intensity reduction in CH over QS in the \ion{Mg}{2} h \& k lines recorded by IRIS, by~\review{considering the line intensity in bins of $\vert B\vert$}. Similar differences in intensities in \ion{C}{2} 1334~{\AA} were also seen by \cite{Upendran_C2} and \cite{Upendran_2022_CHQS} (henceforth {\ppmg}). These authors further found that CHs showed both excess downflows and upflows in the chromospheric lines when the pixels with only downflows or upflows were considered. Furthermore, both the regions showed similar kurtosis in \ion{C}{2} 1334~{\AA} as a function of $\vert B\vert$. These flow velocities and intensities were found to increase with $\vert B\vert$ The authors further demonstrated that the chromospheric~\review{and transition region upflows, as well as the downflows are correlated in the two regions.} For similar chromospheric upflows (downflows), the transition region upflows (downflows) are larger in CHs (QS). Finally, the transition region upflows also \review{exhibit correlation} with chromospheric downflows, which were interpreted as bidirectional flows.

This differentiation between CH and QS, obtained statistically lower in the atmosphere and very strongly in the upper atmosphere, was hypothesized to occur in a unified paradigm in {\ppsi} and {\ppmg}. \review{This paradigm unifies the origin of solar wind in CHs, and enhanced heating in QS, manifested by} interchange reconnection in CH and closed loop reconnection in QS,~\review{respectively}. The reconnection process results in localized, similar levels of heating in both regions,~\review{given a similar amount of magnetic flux in both the regions}. In QS, this heating causes a net local rise in temperature followed by radiative cooling. In the CHs, however, it leads to the expansion of plasma into the solar wind, due to the differences in magnetic field topology. {\ppmg} presents numerous observational and theoretical pieces of evidence in favor of the possible theoretical setup that gives rise to the observations.~\review{However, the authors do not perform numerical experiments to understand the observable implications of their hypothesis, comparing the dynamics in CHs and QS.}

Numerical simulations of dynamics in CHs have been performed in 1~D \citep[e.g.,][]{He2008_SWModelling}, 2~D \citep[e.g.,][]{aiouaz_2005_funnelheating, ding_2010_chjet, ding_2011_chfluxemergence, yang_2013_chjets, yang_2018_chjets} and 3~D \citep[e.g.,][]{Hansteen2010_SWModelling,fang_2014_simulations}. The 2~D or 2.5~D experiments study the dynamic and thermodynamic response of interchange reconnection through either an emergence of flux sheet parameterized at the bottom boundary~\citep{ding_2010_chjet,ding_2011_chfluxemergence}, or through interaction between open flux and closed-loop systems driven by horizontal motion~\citep{yang_2013_chjets,yang_2018_chjets}.

Self-consistent emergence of magnetic flux sheet/tubes has also been studied in many numerical experiments~\citep[][to name a few]{shibata_idealMHD, shibata_resistiveMHD, nozawa_1992_fluxemergence, isobetripathiarchontis, archontis_200_3demergence, archontis_2005_3demergence, galsgaard_2005ApJ_3dreconnection, archontis_2006_3dplasmoids, gaalsgaard_2007A_3dreconnection,nishizuka_2008_simulation,fang_2014_simulations}. Notably, \cite{yokoyama_1996_jetsimulations} perform numerous flux emergence experiments with different coronal magnetic field topologies to simulate emergence and dynamics in different regions of the Sun. For a comprehensive review of flux emergence, see~\cite{cheung_2014_fluxemergence}.  \review{Each of these experiments may or may not include the various thermal sources in their experiments. For instance, \cite{fang_2014_simulations} include thermal conduction in their simulation, but no other thermal sources. However, \cite{shibata_idealMHD, shibata_resistiveMHD, nozawa_1992_fluxemergence,yokoyama_1996_jetsimulations, isobetripathiarchontis, archontis_200_3demergence, archontis_2005_3demergence, galsgaard_2005ApJ_3dreconnection, archontis_2006_3dplasmoids, gaalsgaard_2007A_3dreconnection,nishizuka_2008_simulation} do not include thermal conduction or radiative losses in their experiments. \cite{miyagoshi_2004_thermalconduction}, for instance, perform a similar flux emergence simulation while also incorporating thermal conduction. \cite{ding_2010_chjet, ding_2011_chfluxemergence} consider thermal conduction, and radiative loss, while \cite{yang_2013_chjets, yang_2018_chjets} also consider a heating term in their simulations.}

\cite{Hansteen2010_SWModelling} perform a 3~D flux emergence experiment in a CH-like setup, including thermal conduction, optically thin radiative loss and~\reviewtwo{coronal heating self-consistently caused by the time evolution in the experiment}. They find numerous reconnection events, Ohmic dissipation, and wave processes in the atmosphere. \cite{moreno_2013_ch3djet} perform a 3~D flux emergence experiment with a twisted flux rope in an oblique background field representing a CH topology. Note that the simulations by \cite{moreno_2013_ch3djet} do not incorporate thermal conduction or radiative cooling effects. They report the formation of hot and cool jets and a high-density `wall' around the emerged flux and jets. A similar jet formation was seen also in simulations by \cite{nobrega_2016_cooljet, daniel_2017_simulation, nobrega_2022_nullptreconn}, who studied the formation of jets in CHs. These works included a more realistic treatment of radiation across the solar atmosphere, including spectral synthesis in {\siiv} under non-equilibrium ionization conditions. This heritage of numerical experiments suggests a very complex interaction of open and closed-loop systems that give rise to enhanced flows in different temperatures~\review{in CHs}.

\review{Prior works have dedicated experiments mostly on understanding the dynamics within a CH or any other background magnetic field topology. In this work, we seek to perform a comparative study of the difference in dynamics between CH and QS, by testing out the hypothesis posed in {\ppsi} and {\ppmg}. We seek to understand the differences in plasma response to interchange and closed-loop reconnection. To perform such} comparative study of the interaction of a ``unit" closed flux system with a background open and closed flux system, we perform self-consistent flux emergence ~\review{as the ``closed loop'' that initiates reconnection in different topology. We perform these} experiments in a horizontal coronal background field depicting QS and an oblique background field depicting a CH. We then synthesize synthetic observables in the transition region and coronal spectral lines, computing intensity, velocity, and line widths to understand the spectral signatures of such an interaction. We select several spectral lines observed by IRIS, and to be observed by the Multislit Solar Explorer~\citep[MUSE;][]{muse1,muse2} and SOLAR-C~\citep[EUVST;][]{shimizu2019_euvst,shimizu2020_euvst}. Finally, we perform statistical studies on~\review{the association between properties of these synthetic spectral profiles with the vertical component of magnetic field at z = 0 (i.e. {\bmag} ), and the association between the properties across different spectral lines themselves.}

The remaining paper is organized as follows: In \S~\ref{sec:sim_setup}, we describe our simulation setup, explaining the setup, the different thermodynamic terms, the spectral synthesis, and the computation of moments. Then, we present the results of our simulation in \S~\ref{sec:sim_results}. Finally, we discuss the implication of our work, especially in the context of observations in \S~\ref{sec:discussion}.

%=======================
\section{Simulation setup}\label{sec:sim_setup}
%------------------------------
We aim to self-consistently model the emergence of a flux sheet from the convection zone into the atmosphere and its interaction with a background field throughout its emergence. For this, we consider (i) the flux sheet in the convection zone, (ii) the ambient magnetic field, and (iii) the initial atmosphere. A perturbation in the flux sheet results in its evolution and subsequent interaction with the ambient magnetic field. We solve the MHD equations (Eq.~\ref{eq:MHD}) in a 2.5~D setup, considering all three components of the variables, with the variations only along the horizontal ($x$) and vertical ($z$) directions using the PLUTO code framework~\citep{pluto}\reviewthree{\footnote{The PLUTO userguide may be found at: \url{https://plutocode.ph.unito.it/files/userguide.pdf}}}.  The derivatives and dependence along the $y$ direction (perpendicular to the paper/screen) are ignored. 
Our simulation grid extends from $\sim1.55$~Mm below the photosphere to $\sim82.15$~Mm above, in the z-direction. The grid spacing in the z-direction is uniform from the bottom of the box to 7.75~Mm with 200 cells, above which it increases in a stretched grid~\reviewtwo{in a geometric progression \reviewfour{with a stretching ratio of $\sim1.00725$}} with 350 cells, as also shown in Fig.~\ref{fig:sim_grid} in Appendix.~\ref{sec:app_1}. Along the x-direction, our domain spans $\sim121.21$~Mm, with 520 cells between $\sim40.3$~Mm and $\sim80.6$ Mm while having a logarithmically increasing grid with 15 cells on each side towards both boundaries.~\reviewfive{The boundaries span from $\sim0.3$ Mm to $\sim121.51$ Mm}, The horizontal grid is explained in detail in Appendix.~\ref{sec:app_1}\footnote{Also found in Pg. 33 of the PLUTO userguide. }.

The gas has a specific heat ratio of $\gamma=5/3$, while the gravitational acceleration is taken to be $2.73\times10^4$ cm s$^{-2}$~\citep[following][]{yokoyama_1996_jetsimulations} in the negative z direction. The MHD equations are:

\begin{equation}
    \frac{\partial}{\partial t}
    \begin{pmatrix} \rho \\ \rho\mathbf{v} \\ \mathcal{E} \\ \mathbf{B}\end{pmatrix}+
    \nabla\cdot\begin{pmatrix} \rho\mathbf{v} \\\rho\mathbf{v}\otimes\mathbf{v}- \mathbf{B}\otimes\mathbf{B}/4\pi+\stackrel{\leftrightarrow}{I}p_t\\ (E+p_t)\mathbf{v}-\mathbf{B}(\mathbf{v}\cdot\mathbf{B})/4\pi\\\mathbf{v}\otimes\mathbf{B}-\mathbf{B}\otimes\mathbf{v}\end{pmatrix}^\mathrm{T} = \begin{pmatrix}0 \\ \rho\mathbf{g} \\\rho\mathbf{v}\cdot\mathbf{g}-\nabla\cdot(\eta\mathbf{J}\times\mathbf{B})/ c - \nabla\cdot\mathbf{F}_c-n^2\Lambda(T)+S \\ - \nabla\times(4\pi\eta\mathbf{J})/c\end{pmatrix},
    \label{eq:MHD}
\end{equation}
\noindent where $\rho$ is plasma density, $\mathbf{v}$ is the velocity, $\mathbf{B}$ is the magnetic field \reviewtwo{in the Gaussian system}, $\eta$ is the resistivity, $\mathbf{g}$ is the acceleration due to gravity, $\cdot$ indicates contraction and $\otimes$ showing outer product. All bold quantities denote vectors, while $\stackrel{\leftrightarrow}{I}$ is the unit tensor. The number density ($n$) and plasma density are related as $\rho=n\mu m_u$, where $m_u$ is the atomic mass unit, and $\mu$ is the mean molecular weight ($=0.6724418$). The total pressure $p_t$ is defined as: $p_t = p+B^2/8\pi$, where $p$ is the gas pressure, and the total energy $\mathcal{E}$ is defined as: $$\mathcal{E} = \rho e+\frac{\rho v^2}{2}+\frac{B^2}{8\pi}.$$ 
The specific internal energy $e$ is defined through the equation of state as $\rho e = p/(\gamma - 1)$. The current density \textbf{J} \reviewfive{satisfies Ampere's law given as} $\mathbf{4\pi J} = c\nabla\times\mathbf{B}$ (where $c$ is the speed of light), while $\mathbf{F}_c$ is the thermal conduction flux, $\Lambda(T)$ is the optically thin radiative loss, and $S$ corresponds to background heating term.~\review{While inclusion of viscosity is beyond the scope of this work, it may potentially play a non-negligible role~\citep{Marsch_2006_swcorona,Rempel_2017_viscosity,Chen_2022_viscosity}}.

At the bottom boundary, we use a rigid wall, where the vector components perpendicular to the boundary are reflected, while the tangential components, scalar quantities are symmetrized. %copied from the domain into the boundary zones. 
The top boundary is open \citep[similar to the upper boundary][]{Shibata_1983_outflowboundary}. \reviewfour{We note since that the major dynamics of interest occurs at $z\leq\sim30$ Mm, which is much lower than the top boundary, we do not expect any strong boundary effects~\cite{Shibata_1983_outflowboundary, yokoyama_1996_jetsimulations}. }.  We use a periodic boundary condition in the horizontal direction. We note that $\nabla\cdot B = 0$ is maintained using a constrained transport method~\citep{del2003efficient,londrillo2004divergence,Mignone_2021_CT}. %\reviewfive{We further note that in this scheme the component of B field parallel to the boundary is computed by considering  $\nabla\cdot B = 0$.}

%Note that we have not included the terms involving fluid viscosity since this viscous loss is expected to be less than the resistive, conductive, and radiative loss in the solar corona~\citep[see, for example, the discussion in \S 3.4 in][]{Marsch_2006_swcorona}. 

The normalization of all the physical quantities is performed using the unit density ($\rho_0=1.7\times10^{-7}$ g cm$^{-3}$), length ($L_0=3.1\times10^{7}$ cm) and velocity ($v_0=1.2\times10^{6}$ cm s$^{-1}$). The derived non-dimensionalizing timescale is $\approx26$~s. The pressure and magnetic field are normalized as $\rho_0v_0^2$ and $\sqrt{4\pi\rho_0v_0^2}$, respectively. \reviewfour{From \reviewfive{Ampere's law}, we have the scaling for current $J_{sc} = c\sqrt{4\pi\rho_0v_0^2}/(4\pi L_0) = c\sqrt{\rho_0 v_o^2/4\pi L_0^2}$.} The temperature in \reviewfive{normalized} units is obtained as $T_{C}=p_{C}/\rho_{C}$~\review{where $\rho_C$ and $p_C$ are density and pressure respectively in \reviewfive{the normalized} units}, and is transformed into units of Kelvin through $T = T_{C}\mu m_uv_0^2/k_B$, where $k_B$ is the Boltzmann constant.

% %----------------------------------------- 
% \subsection{Dissipation and redistribution terms} \label{sec:dissip_redistribute}
% %------------------------------------------ 
Our models also include localized resistivity, thermal conduction, and optically thin radiative loss. We consider a localized, anomalous resistivity that depends on the drift velocity, following \cite{sato_1979_fastreconnection,ugai_1992_fastreconnection,yokoyama_1994_fastreconnection,yokoyama_1996_jetsimulations}. The functional form is given by:

\begin{eqnarray}
\eta := &  \left\{
\begin{array}{cc}
0 & \text{ , if $v_{dp}/v_c < 1$}\\
\min\{1,\alpha(v_{dp}/v_c-1)^2\} & \text{, $v_{dp}/v_c\geq1$}
\end{array}\right.
\label{eqn:eta}
\end{eqnarray}
The resistivity is parameterized in terms of $v_{dp}/v_c$, where $v_{dp} = |\mathbf{J}|/\rho$. ~\review{This quantity, though arbitrary, would serve as a proxy for the drift velocity ($|J|/(ne)$) and is not the drift velocity itself}.  $v_c$ is a threshold above which the resistivity effects set in, following \cite{yokoyama_1994_fastreconnection,yokoyama_1996_jetsimulations}. 
\reviewfour{This term is related to drift velocity as: $v_{dp}= |J|/\rho = |J|/(n\mu m_u) = v_d [e/(\mu m_u)]$, where $v_d$ is the drift velocity. The non-dimensionalising factors for  $v_{dp}$ are obtained through its definition as:$$v_{dp,0} =  \frac{J_{sc}}{\rho_0} = \frac{c}{\rho_0}\sqrt{\frac{\rho_0 v_o^2}{4\pi L_0^2}} = c\sqrt{\frac{ v_0^2}{4\pi\rho_0 L_0^2}}.$$ This corresponds to $v_{dp,0} = 7.94\times10^{11}$ g$^{-1/2}$ cm$^{5/2}$ s$^{-2}$. This results in a $v_c = 7.94\times10^{14}$ g$^{-1/2}$ cm$^{5/2}$ s$^{-2}$}, for a value of $10^3$ in code units.  $\alpha=0.01$ in code units,{which is of the same units as $\eta \rightarrow v_0L_0 = 3.7\times10^{13} \mathrm{cm}^2 \mathrm{s}^{-1}$. This results in an $\alpha = 3.7\times10^{11} \mathrm{cm}^2 \mathrm{s}^{-1}$}~\citep[see, for example, a discussion in ][]{yokoyama_1994_fastreconnection,yokoyama_1996_jetsimulations}. The resistivity effects will set in only in regions with high current density (or low density for moderate currents), i.e., typically near current sheets, and will result in a fast, Petschek-like reconnection in the magnetic field setup. %and a $v_{d,0} = 0.00158$ cm s$^{-1}$.

We include anisotropic field-aligned thermal conduction with Runge-Kutta Legendre super time-stepping \citep{vaidya_2017_thcond}. The thermal conduction flux $\mathbf{F}_c$ is defined as: 
\begin{equation}
    \mathbf{F}_c = \kappa_{||}\hat{\mathbf{b}}(\hat{\mathbf{b}}\cdot\nabla T)+\kappa_{\perp}[\nabla T-\hat{\mathbf{b}}(\hat{\mathbf{b}}\cdot\nabla T)],
    \label{eqn:thermalconduction}
\end{equation}
where $\hat{\mathbf{b}}=\mathbf{B}/B$, the unit \review{vector} in the direction of the magnetic field. $\kappa_{||}$ is taken to be Spitzer-type, with $\kappa_{||} = \kappa_0 T^{5/2}$, and $\kappa_0 = 10^{-6}$ erg s$^{-1}$ cm$^{-1}$ K$^{-7/2}$. We ignore the conductivity across the field lines and do not impose any saturation flux in this setup \citep{miyagoshi_2004_thermalconduction}. 

We consider optically thin radiative losses in this work using the CHIANTI database \citep[v10][]{Dere_1997_Chianti,delzanna_2021_chianti}. For computing the radiative losses, we need the characteristic density and a temperature grid. We compute the optically thin radiative loss function over a temperature grid of $\log T/[K]= 4$ to $\log T/[K] = 9$ over 300 points in log space for a number density of  $10^{11}$~cm$^{-3}$ and use coronal abundances \citep{fludra_1999_coronalabundance}. The radiative loss is however given by $n^2\Lambda(T)$, where $\Lambda(T)$ is the radiative loss function as also shown in Eq.~\ref{eq:MHD}. In our simulation, we make the radiative loss zero if the temperature falls below $\approx85000$ K or if the number density exceeds $10^{13}$~cm$^{-3}$, which is typical of chromosphere or lower transition region, and the optically thin radiative cooling formalism fails in these conditions. Furthermore, we do not have any radiative loss in the convection zone \citep{takasao_2013_convectRadloss}.

Note that in these simulations, we do not have a self-consistently generated heating in the corona like some of the other models \citep[see, e.g.,][]{Hansteen2010_SWModelling}. Hence, we add a background heating term ($S$ in Eq.~\ref{eq:MHD}),~\reviewtwo{constant in time,} to compensate for the radiative cooling. At t=0, we ensure there is no net dissipation or heating, and define the heating term at each grid point to be the same value as the radiative cooling at that point, following \citep{roussev_2001_bhgeating}. This heating term is defined to compensate only for the cooling term when there are no dynamics in the system.

The radiative loss function has a complicated dependence on the temperature, consisting of a general reduction with an increase in temperature and containing localized bumps~\reviewtwo{for the temperature range in consideration}. If due to a numerical error, the cooling term becomes slightly smaller than the heating term, the grid point will be at a higher temperature. However, at the next iteration, this results in lower radiative cooling, and causes more heating. On the other hand, a slightly higher cooling can result in a runaway cooling of the system. This instability depends on the behavior of the heating and cooling terms with temperature \citep[see][for details]{parker_1953_thermalinstab,shimojo_2001_radjets}. Thus, even after the inclusion of the heating term, numerical errors may build up over time, and result in runaway heating or cooling of the system. To mitigate this, we impose a numerical floor on the ``net heating'' term, i.e., if $|S-n^2\Lambda|<10^{-3}S$, there is no heating or cooling in that grid cell, mitigating the effect of numerical instability.
%-------------------------------------------
\subsection{Initial condition and background field}~\label{sec:init_cond}
%-------------------------------------------
The atmosphere is assumed to be initially in magnetohydrostatic equilibrium. To define the atmosphere, we first specify the temperature profile of the gas. The atmosphere consists of three parts, following \cite{yokoyama_1996_jetsimulations}: a convection zone, a photosphere~\reviewtwo{/chromosphere}, and a corona. For $z\geq0$, the temperature is defined as:

\begin{equation}
    T(z) = T_{phot}+ \frac{(T_{cor}-T_{phot})}{2}\left(\tanh\left(\frac{z-z_{tr}}{w_{tr}}\right)+1\right),
    \label{eqn:temp_atm}
\end{equation}
while for $z<0$, it is defined as:
\begin{equation}
    T(z) = T_{phot} -  a\left| \frac{dT}{dz}_{ad} \right| z.
    \label{eqn:temp_ad}
\end{equation}
\noindent Here, $T_{phot}$ is the photosphere/chromospheric temperature ($=17319$ K), $T_{cor}$ is the coronal temperature ($=1.7319\times10^6$ K), $z_{tr}$ is the height , $w_{tr}$ is the width of the transition region. $|dT/dz|=1-1/\gamma$ is the adiabatic temperature gradient. Under this setup, for $a>1$, the layer becomes convectively unstable \citep{nozawa_1992_fluxemergence}. Following \cite{yokoyama_1996_jetsimulations}, we have used $a=2$ in our simulations. 

As alluded to earlier, we have two models with different background fields, one with a horizontal background field (\modqs) and another with an oblique background field (\modch) mimicking QS and CH topologies, respectively. The height of the transition region is known to be different in QS and CH, which is incorporated in these simulations through the height of the transition region $z_{tr}$. This value is fixed at $\sim2480$ km in \modqs, while it is fixed at $\sim4500$ km in \modch, following \cite{Tian_TRheight}. The width of the transition region is $\sim155$ km in both simulations.

In \modqs, we consider a formulation with the flux sheet in the convection zone and a horizontal field in the corona, following \cite{miyagoshi_2004_thermalconduction}. We define the magnetic field as:

\begin{equation}
    B(z) = \sqrt{\frac{8\pi p(z)}{\beta(z)}},
    \label{eqn:bz}
\end{equation}

\noindent where $\beta(z)$ is the plasma beta and is defined as: 

\begin{equation}
    \frac{1}{\beta(z)} = \frac{1}{\beta_{fs}(z)}+\frac{1}{\beta_{bg}(z)},
    \label{eqn:beta}
\end{equation}

\noindent where $\beta_{fs}(z)$ specifies the flux sheet, and $\beta_{bg}(z)$ specifies the background in the {\modqs} case. The flux sheet is defined as:
\begin{equation}
    \frac{1}{\beta_{fs}(z)} = \frac{1}{4\beta_{fs0}}\left(1+\tanh\left(\frac{z-z_{fsL}}{w_{fsL}}\right)\right)\left(1-\tanh\left(\frac{z-z_{fsU}}{w_{fsU}}\right)\right),
    \label{eqn:fluxsheet}
\end{equation}

\noindent where $\beta_{fs0}$ is the plasma beta at the center of the flux sheet (=4.0), $z_{fsL}$ is the lower end (=-1240 km), and $z_{fsU}$ (=-620 km) is the upper end of the flux sheet, while $w_{fs}$ (=155 km) determines the length scale of field increase and reduction from the background medium to flux sheet. The plasma beta at the center of the flux sheet is a crucial factor in determining the emergence time scale of the flux sheet.

%------------------------
\begin{figure}[ht!]
    \centering
    \includegraphics[width=0.48\linewidth]{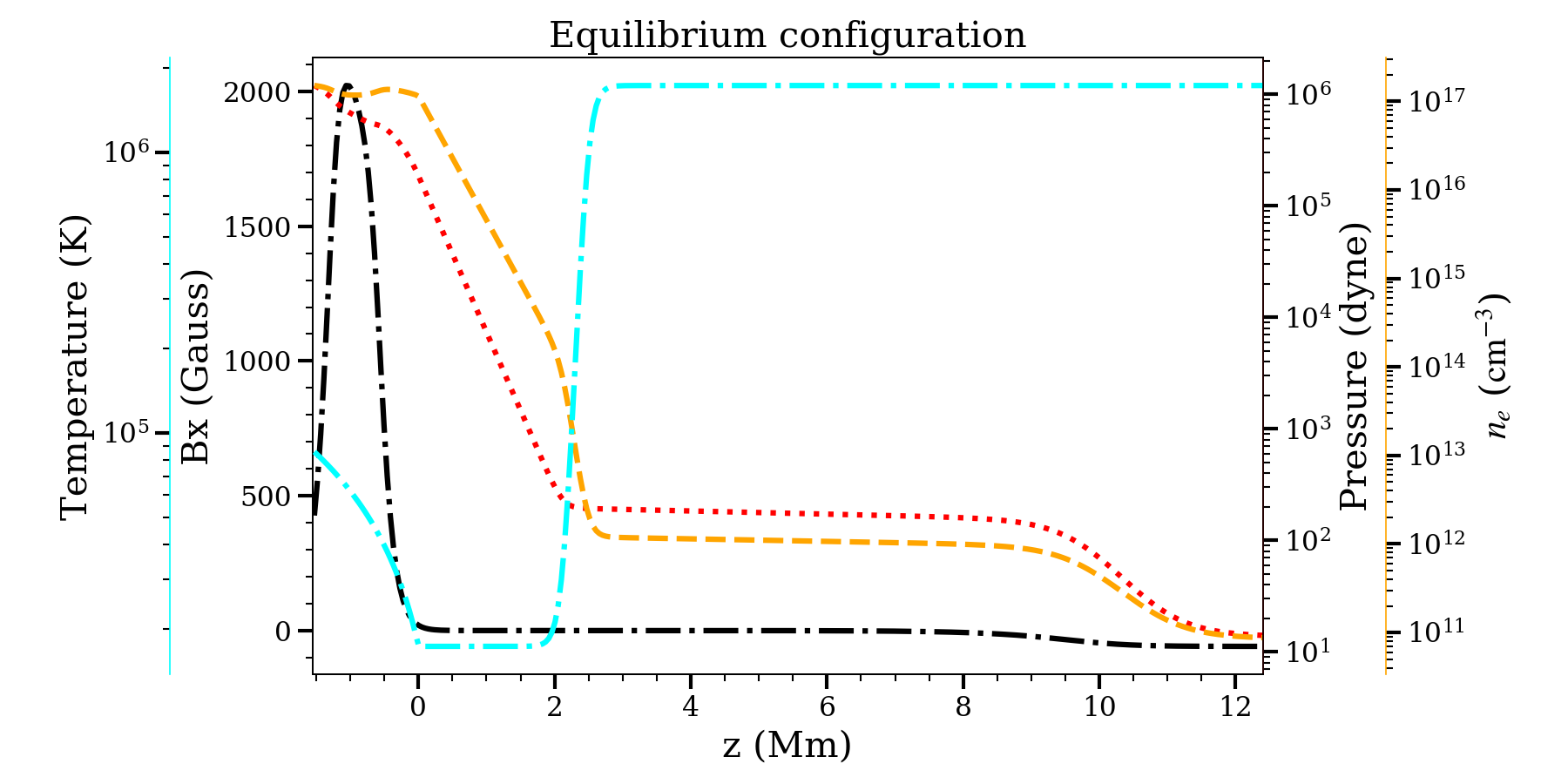}
    \includegraphics[width=0.48\linewidth]{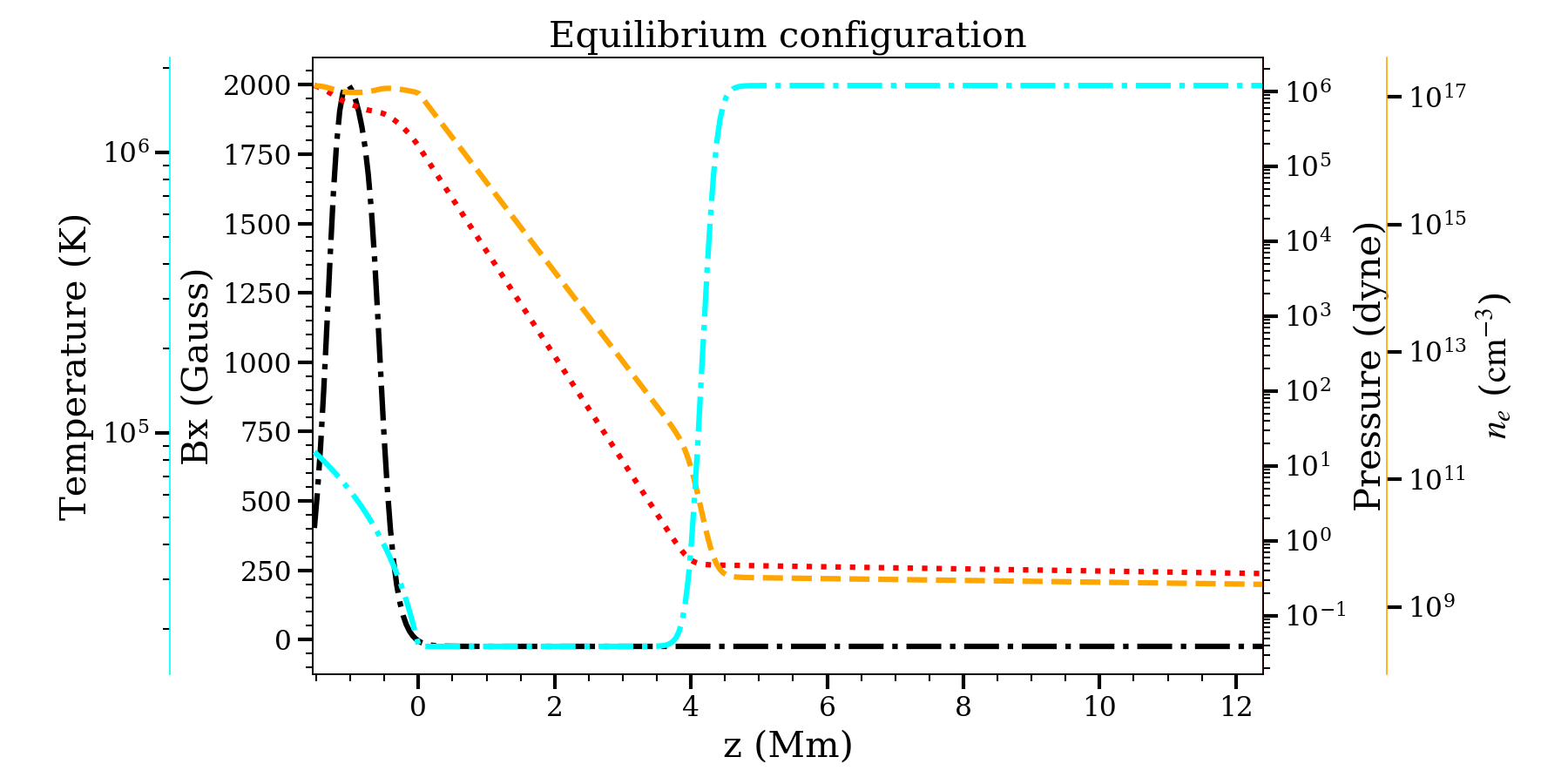}
    \caption{Initial configuration of the system in {\modqs} (left) and {\modch} (right). We show the number density $n$ (yellow dashed line), Temperature (cyan dot-dashed line), horizontal magnetic field $B_x$ (black dot-dashed line), and gas pressure (red-dotted line) along the center of the box. }
    \label{fig:sim_init}
\end{figure}
%------------------------

In {\modqs}, we define the coronal magnetic field as:
\begin{equation}
    \frac{1}{\beta_{bg}(z)} = \frac{1}{2\beta_{bg0}}\left(1+\tanh\left(\frac{z-z_{cor}}{w_{cor}}\right)\right),
    \label{eqn:qsbg}
\end{equation}

\noindent where the coronal field starts from $z_{cor}$ (=\review{2480} km), rises with width $w_{cor}$ (=852.5 km), and has a plasma beta of $\beta_{bg0}=0.1$. However, note that to force reconnection between the emerging flux and the background field, the coronal field is oriented opposite to the field in the convection zone. 

We then determine the initial variables by solving the magnetohydrostatic equilibrium equation: 

\begin{equation}
    \frac{d}{dz}\left(p(z)+\frac{B^2(z)}{8\pi}\right) +\rho(z)g = 0.
    \label{eqn:static}
\end{equation}

In {\modch}, we do not use the formulation of the coronal field from Eq.~\ref{eqn:qsbg}. We \reviewtwo{once again} consider the flux sheet in the convection zone and impose a uniform, time-independent oblique field \reviewtwo{as a background field~\citep{powell_1994_bgfieldsplit}} across the whole box, with a strength of $\approx40$ Gauss, and oriented at $3\pi/4$ from the x-axis. \reviewthree{We note that the boundary conditions are applied only on the changing field}. \reviewtwo{Note also that we have a ``hot plate'' at the top boundary in {\modch}, which seeks to maintain the temperature of the top boundary at $10^6$ K~\citep[see, for example][]{Leenaarts_hotplatetopboundary, Ijima_2015_hotplate, daniel_2016_hotplate}. Thus, the pressure is set by the outflow condition of density and the fixed temperature. } Furthermore, in this case, the application of a hot plate top boundary also causes background temperature changes to be redistributed by thermal conduction, which vastly reduces the numerical discrepancy between the heating and cooling terms, preventing runaway heating or cooling of the simulation box.

The initial condition for different system parameters along a vertical column for {\modqs} and {\modch} are depicted in Fig.~\ref{fig:sim_init}.

With the system defined, we then perform a perturbation of the flux sheet in the vertical velocity as:

\begin{equation}
    V_z=A\cos\left(2\pi\frac{x-X_{range}/2}{\lambda_p}\right),
    \label{eqn:perturb}
\end{equation}

\noindent where the perturbation is performed in the middle of the flux sheet, with an amplitude of $A=0.6$ km s$^{-1}$, and a $\lambda_p=6200$km, and $X_{range}$ is the horizontal expanse of the box ($\sim121.21$~Mm). This wavelength is almost the most unstable wavelength for linear Parker instability. Note that the perturbation is performed only within $X_{range}/2-\lambda_p/4<x<X_{range}/2+\lambda_p/4$ and $z_{fsL}<z<z_{fsU}$.

%-------------------
\subsection{Spectral synthesis} \label{sec:specsynth}
%------------------------
Given the model dynamics, we also wish to study the response in different spectral lines. Hence, we perform synthesis of optically thin spectral lines that form in the corona, and the transition region. The list of lines we synthesize, \reviewfour{their rest wavelengths, and their approximate formation temperatures} are provided in Table~\ref{tab:linelist}.%These lines are either observed by IRIS or will be observed by MUSE and SOLAR-C EUVST.

We assume the lines are formed in the optically thin regime and are in ionization equilibrium. In the MHD approximation with completely ionized hydrogen, we have the intensity per unit frequency in a given line defined as:
\begin{equation}
    I_\nu = \int_{h} n^2 \phi(\nu) \mathrm{G(T,n)} dh,
    \label{eqn:I}
\end{equation}
\noindent$\mathrm{G(T,n)}$ is the contribution function of a spectral line, $\phi(\nu)$ is the spectral profile as a function of frequency, and $h$ is the column depth along the line-of-sight (LOS). The spectral profile is defined as a Gaussian along with \reviewtwo{thermal} broadening as:
\begin{equation}
    \phi(\nu) = \frac{1}{\sqrt{\pi}\Delta\nu_{th}}\mathrm{exp}\left[-\left(\frac{\Delta\nu-\nu_0 v_{\mathrm{LOS}}/c}{\Delta\nu_{th}}\right)^2\right],
    \label{eqn:gaussprofile}
\end{equation}
where $\nu$ is the frequency range, $\Delta\nu = \nu - \nu_0$, the difference from rest frequency ($\nu_0$), $c$ is the speed of light, $v_{\mathrm{LOS}}$ is the line of sight velocity, and $\Delta\nu_{th}$ is the thermal width of the line defined as:
\begin{equation}
    \Delta\nu_{th} = \frac{\nu_0}{c}\sqrt{\frac{2k_BT}{M_{ion}}},
    \label{eqn:thermalwidth}
\end{equation}
where $M_{ion}$ is the mass of the ion in consideration.

We note that in some cases we do not perform any integration along the line of sight. In such cases, we define the \reviewfive{emission contribution to intensity per line of sight element} as a function of two spatial dimensions as:
\begin{equation}
    E_{\nu,x,z} = n^2 \phi(\nu) \mathrm{G(T,n)}
    \label{eqn:i_nu_0}
\end{equation}
\reviewfour{In such cases, we only consider the line profile at each point.}
We consider a velocity grid of $\pm100$ km s$^{-1}$ at a resolution of $0.2$ km s$^{-1}$, and convert it to frequency to obtain $\nu$. We compute the contribution function $\mathrm{G(T,n)}$~\citep[using CHIANTIPy:][]{dere2013chiantipy,barnes2017chiantipy}\review{\footnote{Using \texttt{ChiantiPy.core.ion.spectrum()}}} for a temperature grid of $\mathrm{\log T/[K]} = [4.5,6.5]$ with a spacing of $\Delta\mathrm{\log T/[K]} = 0.1$, and a density grid of $\mathrm{\log n/[cm^{-3}]} = [8,18]$ with a spacing of $\Delta \mathrm{\log n/[cm^{-3}]} = 0.1$. For each grid point of our simulation, we perform a grid search in temperature and density to find the associated $\mathrm{G(T,n)}$, and then compute the intensity. 
%----------
\begin{table}[ht!]
    \centering
    \begin{tabular}{c|c|c }
    \hline 
     Line & $\lambda$ (~{\AA}) & $\mathrm{\max [\log T/[K]] }$ \\
     \hline 
     \hline
     \fexv &  284.1630 & 6.4\\
     \feix &  171.0730 & 5.95 \\
     \neviii & 770.4280 & 5.8 \\
     \ov & 629.7320 & 5.35 \\
     \oiv & 1401.1630 & 5.1 \\
     \siiv & 1393.7550 & 4.85\\
     \hline 
    \end{tabular}
    \caption{List of spectral line forward modeled in this work, with their rest wavelength and approximate formation temperatures. }
    \label{tab:linelist}
\end{table}

%-------------------
\subsection{Calculating moments} \label{sec:specsmoments}
%------------------------
To compute physical properties like intensity, velocity, and line width, we compute the moments associated with the spectral lines \reviewfour{integrated over the line of sight, as defined in Eq.~\ref{eqn:I}}. Consider a spectral line at each grid position as $I_{v_D}(v_D)$, as a function of velocity axis $v_D$ (rest frequency $\nu_0$ corresponds to 0 velocity~\reviewtwo{, and $v_D = (\nu/\nu_0 - 1)*c$} ). \reviewfour{We define the net intensity within the line at each grid point and} calculate the moments as:
\begin{subequations}
    \begin{gather}
    \begin{align}
      I &= \int_{v_D}~I_{v_D}(v_D)~dv_D \\
      v_{LOS,p} &= \frac{1}{I}~\int_{v_D}~I_{v_D}(v_D)~v_D~dv_D  \\
      W &= \sqrt{\frac{1}{I}~\int_{v_D}~I_{v_D}(v_D)~(v_{LOS} - v_D)^2 dv_D},
    \end{align}
    \end{gather}
    \label{eqn:moments}
\end{subequations}
\noindent where $I$ is the \reviewfive{intensity within the line}, $v_{LOS,p}$ corresponds to the Doppler shift~\reviewthree{at each pixel} and $W$ to the line width. \reviewfour{If we consider the quantity $E_{\nu,x,z}$ defined in Eq.~\ref{eqn:i_nu_0}, and integrate across all frequencies, we define this quantity (in units of erg cm$^{-3}$ s$^{-1}$ sr$^{-1}$) as $E_{x,z}$ such that:}
\begin{equation}
    E_{x,z} =  \int_{\nu}~I_{\nu,0}(\nu, x, z)~d\nu
    \label{eqn:i_0}
\end{equation}
\reviewfour{We use $E_{x,z}$ to study the spatial distribution of emission. The moments defined in Eq.~\ref{eqn:moments}} are analyzed to understand the observable properties of this flux emergence interaction. Because of our definition of velocity (outward along z is positive), any upflows along z will result in a reduction of emission frequency in Eq.~\ref{eqn:gaussprofile}. Hence, upflows towards the observer at the top of the box, along $z$, would correspond to positive shifts in velocity, unlike the conventional definition adopted by observers.
%-====================================
\section{Simulation results}~\label{sec:sim_results}

We now present the results for the two simulations namely the QS simulation {\modqs} and the CH simulation {\modch}.  In both cases, we first present the time evolution of the density and temperature of the system. Then, we showcase the spectral response and study the integration along the vertical direction. We also perform a statistical comparison of the intensity, velocity, and line widths as a function of {\bmag}. Finally, we study the relation between flows inferred from different spectral lines and investigate the coupling between different observables.
%-----------------
\begin{figure}[t!]
    \centering
    \includegraphics[width=\textwidth]{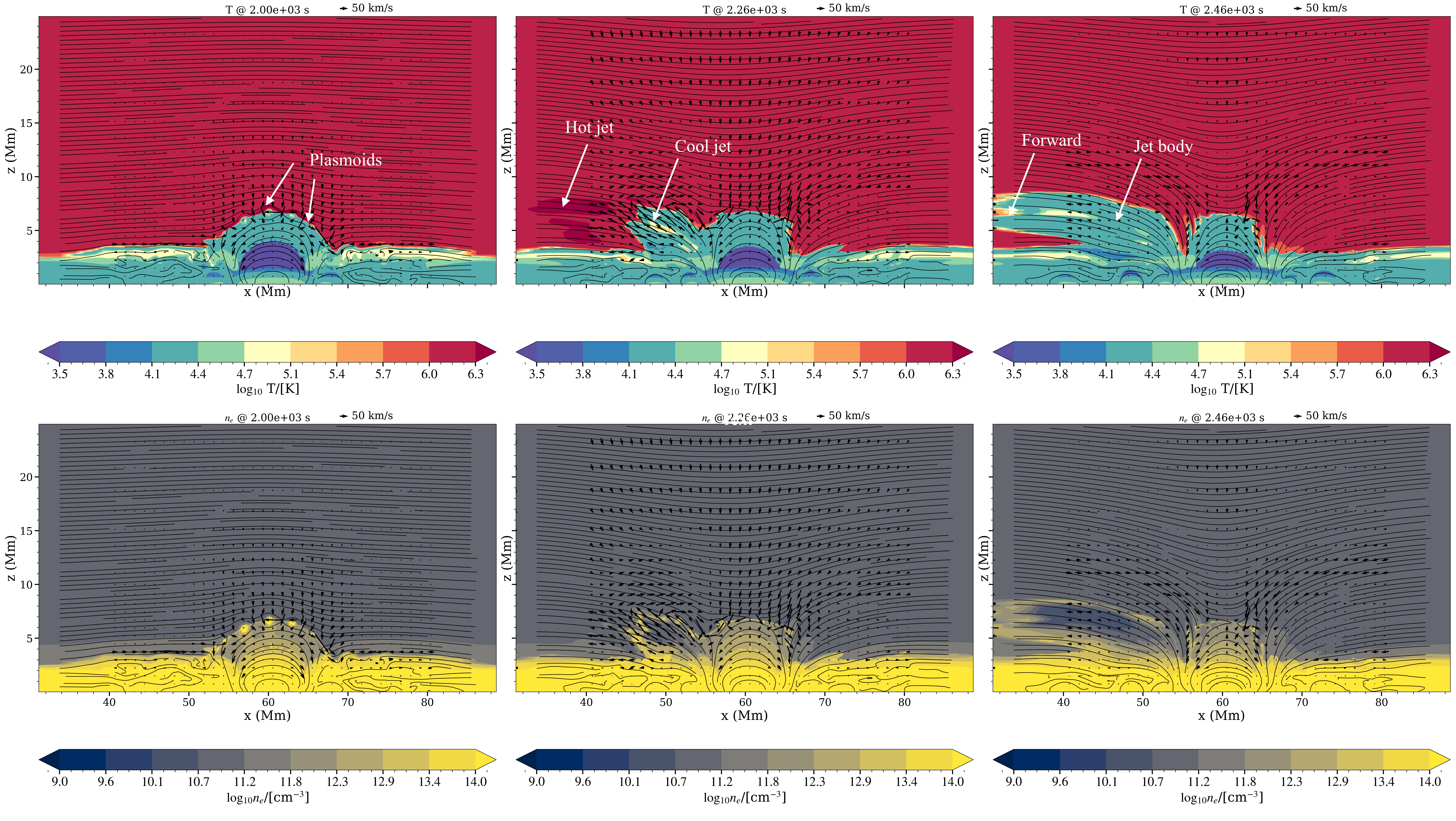}
    \caption{Temperature (top row) and number density (bottom row) evolution in {\modqs}. The three timestamps correspond to the onset of reconnection, the resultant jet formation, and the ``steady state'' jet at a later time. The black contours represent magnetic field lines. The arrows depict velocity flows, with the unit size of 50 $\mathrm{km}\mathrm{s}^{-1}$ as depicted at the top of each plot.  We also refer the reader to animation QS1, depicting the evolution of the temperature and density respectively, \reviewthree{of which three timestamps are presented in-text}. \reviewtwo{The animation follows through the evolution of flux sheet and the resultant jet formation, of which the most important dynamics of interest are displayed in the static image.}  }    
    \label{fig:qs_density_temp}
\end{figure}
%---------------

%===********==============================================================
%=========================================================================
\subsection{{\modqs}: Results} ~\label{sec:qs_results}
%===********==============================================================
%=========================================================================

%-====================================
\subsubsection{Dynamics} ~\label{sec:qs_dynamics}
%-====================================
We present the evolution of temperature and density in {\modqs} in Fig.~\ref{fig:qs_density_temp}. The emerging flux sheet undergoes reconnection with the ambient magnetic field, resulting in the formation of plasmoids, which are expelled on either side of the reconnection region (at $2000$~s). These plasmoids collide with the ambient atmosphere and result in the formation of ~\reviewtwo{hot and cool }jets. \review{Near the forward edge of the cool jet, where the plasmoids interact with the ambient atmosphere, we find the formation of ``hot jets'' (at $2260$ s).} The plasmoids typically have a temperature of  $\approx3\times10^4$ K and densities of $\approx10^{13}$ cm$^{-3}$. They travel outward with a typical speed of $\approx50$~km/s. \review{The dense, cool jets} are at temperatures of $\approx2-5\times10^4$ K, traveling outward with a speed of $\approx50$ km/s (for example, between x = 45 and 50 Mm and z = 3 and 8 Mm at $2460$~s). The forward edge of the jet corresponding to the region of interaction of plasmoids with the ambient atmosphere has high densities ($\approx10^{11}-10^{13}$ cm$^{-3}$), while the body of the jet has low densities ($\approx10^{10}$ cm$^{-3}$), as seen at $2260$ s and $2460$~s. Due to its lower density, the hot jet does not experience any strong cooling. The hot jet has a characteristic density of $\approx10^{10}$ cm$^{-3}$, and temperature of the order of $\approx10^6$ K. Finally, the jets in this experiment are observed to be very low-lying ($\approx4-10$)~Mm, and almost horizontal at later times. The dynamics observed here are similar to those by~\cite{yokoyama_1996_jetsimulations,miyagoshi_2004_thermalconduction}. The dense jets are reminiscent of the slingshot-like motion of plasma from \cite{yokoyama_1996_jetsimulations}. 

%-====================================
\subsubsection{Synthetic observables}~\label{sec:qs_spectrum}
%-====================================
\begin{figure}[t!]
    \centering
    \includegraphics[width=\textwidth]{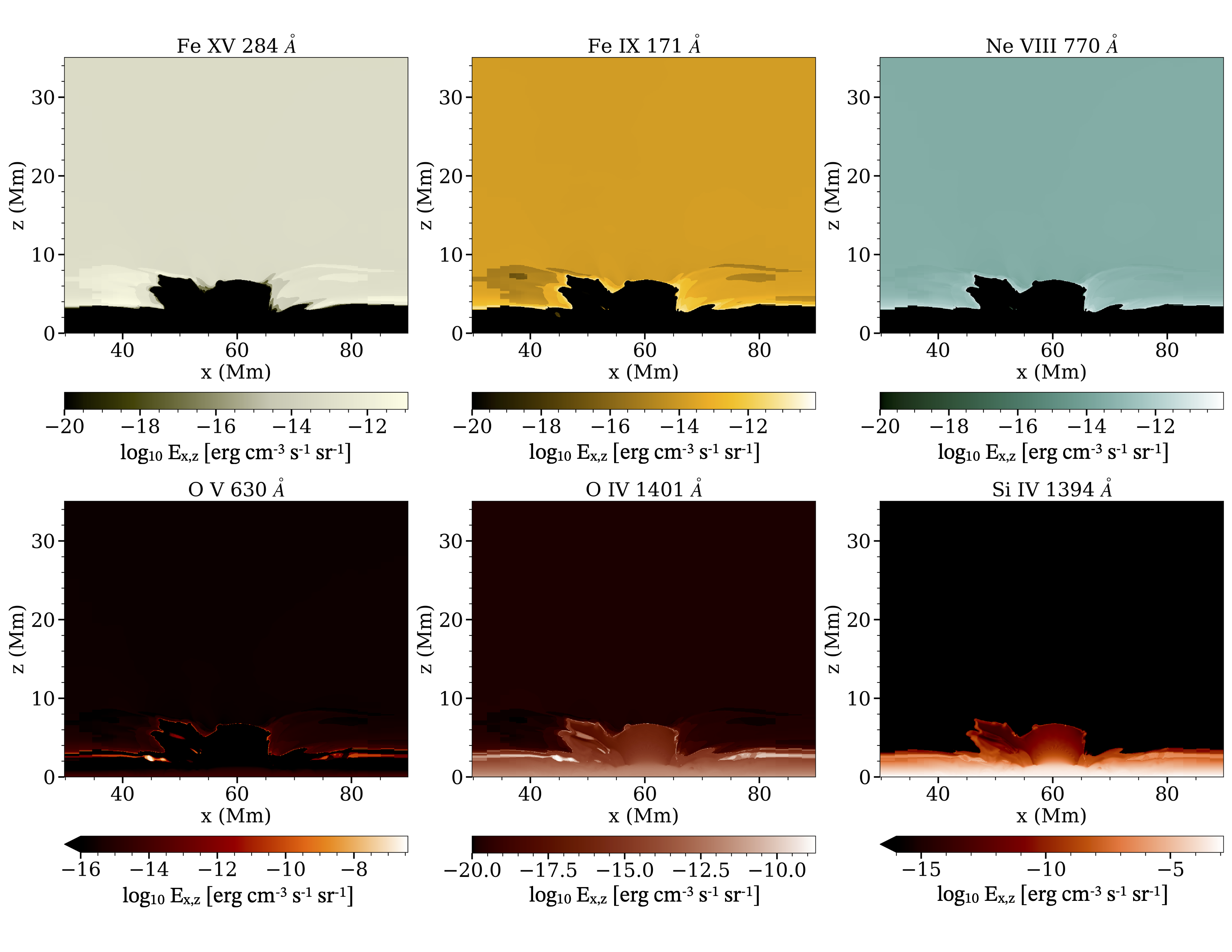}
    \caption{Intensity in \ion{Fe}{15}~284~{\AA}, \ion{Fe}{9}~171 ~{\AA}, \ion{Ne}{8}~770~{\AA}, \ion{O}{5}~630~{\AA}, \ion{O}{4}~1401~{\AA}, and \ion{Si}{4}~1394~{\AA} for {\modqs} snapshot at t = 2260~s displayed in the middle panels in Fig.~\ref{fig:qs_density_temp}. We also refer the reader to animation QS2,~\reviewtwo{where the spectral response corresponding to QS1 is displayed. The static image displays the most important dynamics of interest.} } 
    \label{fig:qs_lines}
\end{figure}
%-----------------------------

We perform spectral synthesis for multiple lines for the snapshot at t = 2260~s shown in the middle panels in Fig.~\ref{fig:qs_density_temp} and compute the moments of various spectral lines. In Fig.~\ref{fig:qs_lines}, we plot the intensity maps obtained from spectral synthesis for {\fexv}~284~{\AA}, {\feix}~171~{\AA}, {\neviii}~770~{\AA}, {\ov}~630~{\AA}, {\oiv}~1401~{\AA}, and {\siiv}~1394~{\AA} as labeled. 

We note the following salient features of the spectral synthesis. First, a gradual change in the structures that gives rise to the emission in different lines is seen across the corona to the lower transition region. \review{For instance, the emission of cool plasma in the low-lying loops and edges of the cool jet are seen in {\siiv}. This changes to an increasing contribution of hotter plasma around the emerged loop in {\oiv}, and mostly probes the hot plasma in other lines.} The cool jet shows signatures in {\siiv} and {\oiv}, but not in hotter lines. \review{The forward edge of the jet} shows signatures in {\neviii} and {\feix}, \review{while the hot jet shows signatures in {\fexv}}. The {\siiv} \review{emission in the cool jet, however, is rather weak ($\approx10^{-8}$ erg cm$^{-3}$ s$^{-1}$ sr$^{-1}$), when compared to the emission from structures lower in the atmosphere ($\approx10^{-3}$ erg cm$^{-3}$ s$^{-1}$ sr$^{-1}$).}

%-====================================
\subsubsection{Space-time rasters}~\label{sec:qs_spacetime}
\begin{figure}[h!]
    \centering
    \includegraphics[width=\textwidth]{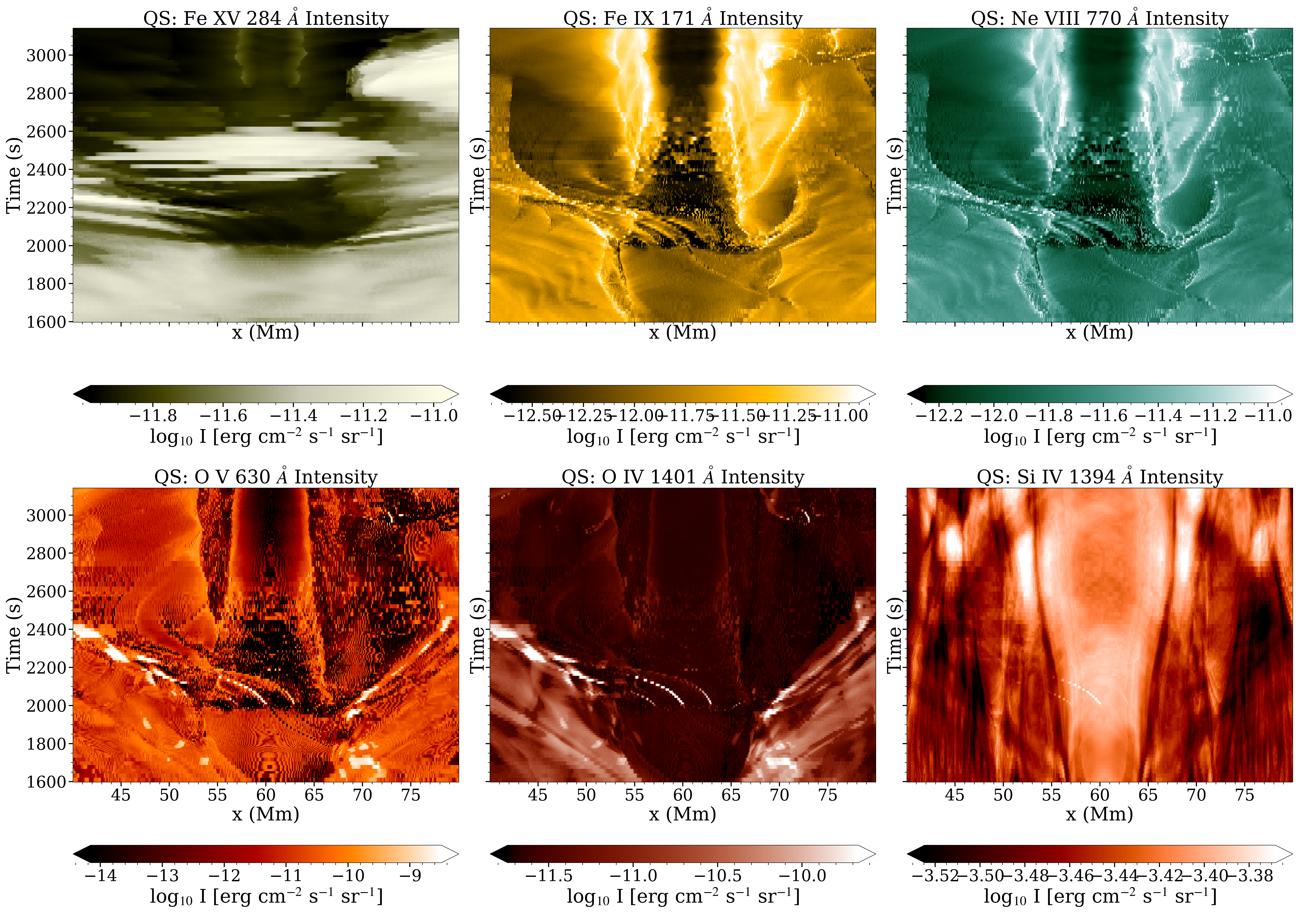}
    \caption{Variation of intensity in different spectral lines, with the line of sight along the vertical direction, and displayed as a function of the horizontal axis and time for {\modqs}. See in-text for details.}
    % {\siiv} is dominated by the mission of low-lying loops, and has a mild signature of plasmoids as moving dots. {\oiv} and {\ov} show strong signatures of propagating plasmoids as bright spots. These signatures are seen mildly as moving streaks in {\neviii} and {\feix}. The temperature redistribution occurring after the jet is seen as parallel bright features in {\neviii} and {\feix}.}
    \label{fig:qs_netflux_xt}
\end{figure}

\begin{figure}[h!]
    \centering
    \includegraphics[width=\textwidth]{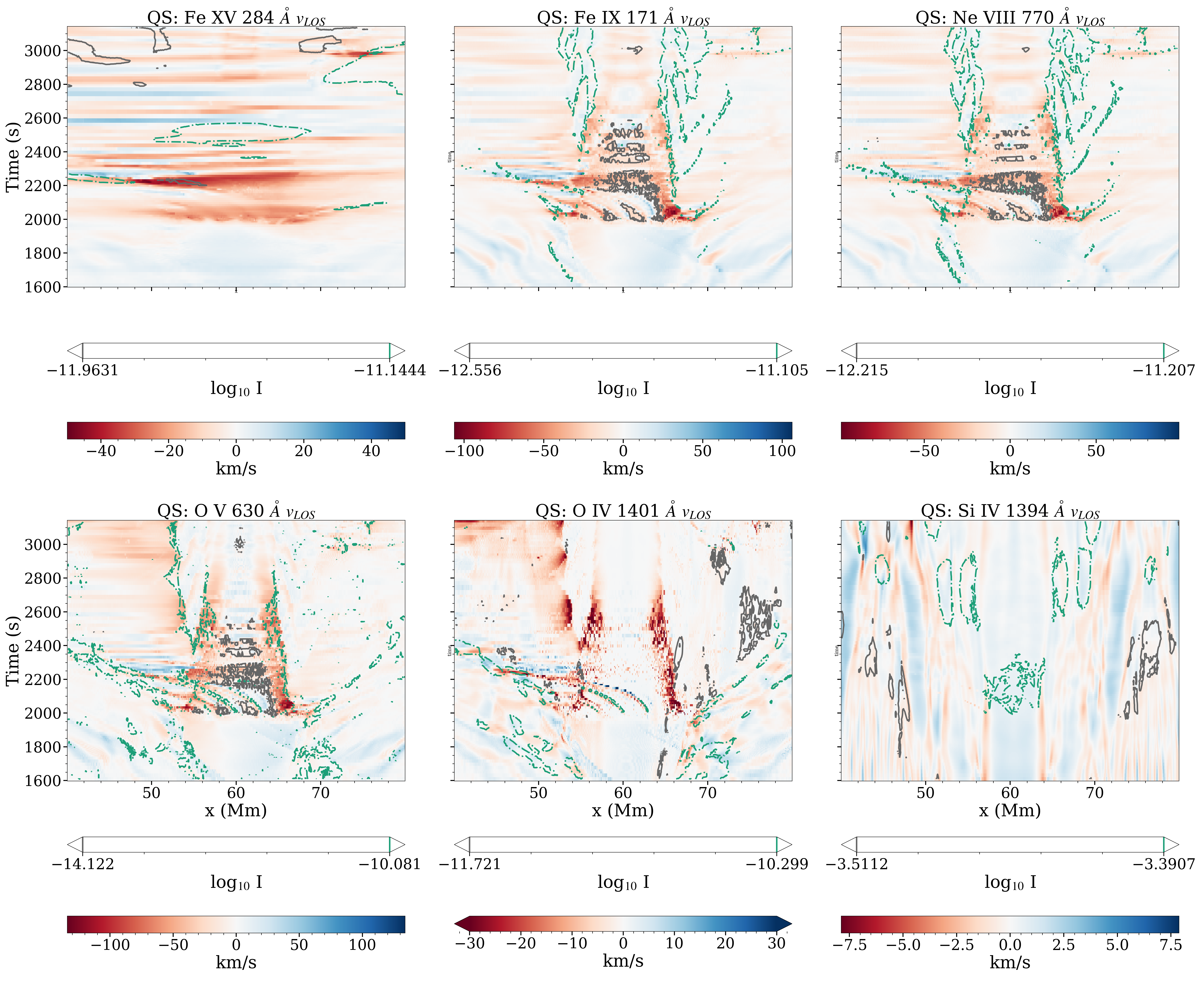}
    \caption{Same as Fig.~\ref{fig:qs_netflux_xt}, but for velocity of {\modqs}. The blue color corresponds to upflows \textbf{towards} the observer, while the red color to the flows \textbf{away} from the observer. Green dot-dashed contours show regions with high intensity while black solid ones depict low intensity. Notice the similarity in the velocity structure of {\ov}, {\neviii}, and {\feix}. Notice also the mild plasmoid signature seen in {\oiv}, {\ov}, {\neviii}, and {\feix}. }
    \label{fig:qs_velocity_xt} 
\end{figure}

\begin{figure}[h!]
    \centering
    \includegraphics[width=\textwidth]{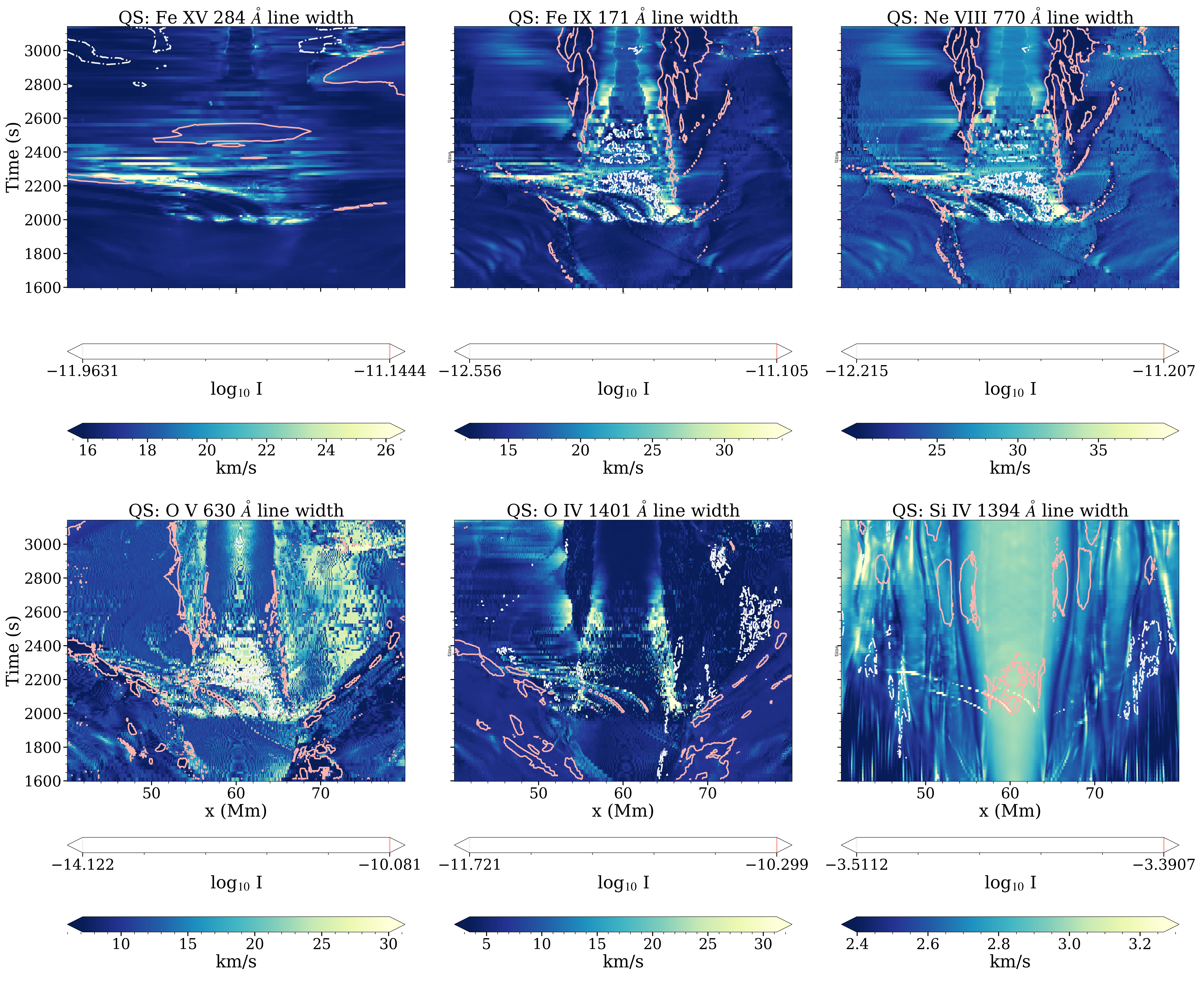}
    \caption{Same as Fig.~\ref{fig:qs_netflux_xt}, but for line width from {\modqs}. Pink solid contours show regions with high intensity while white dot-dashed ones depict low intensity. The plasmoids show signatures in {\feix}, {\neviii}, {\ov}, {\oiv}, and {\siiv}. Furthermore, the strongest line widths show a strong correspondence with the strongest velocities of Fig.~\ref{fig:qs_velocity_xt}. }
    \label{fig:qs_width_xt} 
\end{figure}

%-====================================
We now study the evolution of intensity, velocity, and line width of different lines for {\modqs} by integrating the spectrum along the vertical ($z$) axis. The evolution is shown from a couple of snapshots just before the onset of reconnection to the end of the simulation.  We show the space-time plots for the intensity (Fig.~\ref{fig:qs_netflux_xt}), velocity (Fig.~\ref{fig:qs_velocity_xt}), and line width (Fig.~\ref{fig:qs_width_xt}) in different spectral lines. In each panel of the three plots, we have the horizontal dimension as x-axis and time on the vertical axis. 

%-====================================
%-====================================
%-====================================
%-====================================
%-====================================
%-====================================

Figure~\ref{fig:qs_netflux_xt} demonstrates the evolution of features as seen from the top of the box. {\siiv} emission is the strongest amongst all the lines and is mainly dominated by the ``low-lying loop'' emission arising from the emergent flux sheet as also seen in Fig.~\ref{fig:qs_lines}. Plasmoids are seen here as small dots evolving from (x,t) $\approx$ (60 Mm, 2050 s), and move outward till t $\approx2200$ s. The enhancements in intensity seen at various locations from t $\approx2600$ and $\approx2800$ s correspond to emergence and reconnection dynamics of \review{the secondary loops that arise at later times}. The plasmoid signatures are very clearly seen in {\oiv} and {\ov} as propagating bright spots. These plasmoids also show signatures in {\neviii} and {\feix}. In these hotter lines, the edges of plasmoids are seen as streaks of moving structures co-temporal with the plasmoids in {\oiv} and {\ov}.

In {\neviii} and {\feix}, we observe strong intensity enhancements near x = $55$ and $65$ Mm from t = $2400$~s that persist for more than 800~s, till the end of the simulation. \review{These persistent enhancements arise due to the interaction of multiple processes (we refer to animation QS1 and QS2). First, we find the return flow of hot plasma along the edge (towards the cooler side) of the jet structures. The in-flowing hot plasma collides with the photospheric plasma and is then reflected upwards with lower temperature along the reconnected field line. Furthermore, we also have the reconnection outflow coming in from the emerged loop, driving plasma flows. The temperature gradient also results in enhanced thermal conduction in those regions. The continuous feed-in of plasma and reflection, along with thermal conduction gives rise to  to these persistent structures bright in {\feix} and {\neviii}.} In {\fexv}, we observe an intensity enhancement localized between t = $2400$ and $2600$ s. We note this as a localized enhancement in temperature (Fig.~\ref{fig:qs_density_temp} at 2580 s), arising due to enhanced local heating or reduced radiative losses. We hypothesize that the reconnection process results in flux being pulled in, perpendicular to the current sheet. This causes low-density plasma to be pulled lower into the atmosphere, resulting in reduced radiative loss at those grid points. Since the background heating is kept constant at each grid point, we find an enhancement in temperature, resulting in the localized enhancement in {\fexv} intensity. However, this effect goes away as the system relaxes after the initial explosive process. 

In Fig.~\ref{fig:qs_velocity_xt}, we depict the line of sight velocity \reviewthree{for each spectral line}, with downflows moving away from the observer in red color and upflows moving towards the observer in blue. Intensity contours are plotted on top, with the black solid lines corresponding to low intensity (10 percentile of intensity distribution) and the green dot-dashed ones corresponding to high intensity (95 percentile for intensity distribution). First, we find a clear gradation of velocity structures from {\siiv} to other lines. {\oiv}, {\ov}, {\neviii}, and {\feix} show very similar velocity structures, which is not seen in {\fexv}. The general upflow seen in {\siiv} near x = $60$ Mm corresponds to the plasma flow due to the rising loop. The strongest of velocities seen here do not seem to have a strong correspondence with the brightest or the darkest structures seen in Fig.~\ref{fig:qs_netflux_xt}, including a near absence of plasmoid signatures. In contrast, the plasmoids show signatures in velocity as upflowing and downflowing blobs in all hotter lines except {\fexv}, seen through the association between intensity contours and velocity signatures. 

We also find an association between the edges of the strong intensity with velocity, as seen near x = $56$ Mm and $65$ Mm in these lines. The excess downflows, for example at x = $66$ Mm and between t = $2050$ and $2100$ s in {\ov}, {\neviii}, and {\feix}, are seen at the boundaries of the low and high intensity structures. The evacuated loop top is seen as low intensity, while the strong downflow is a signature of downflows along the loop toward the footpoints. From {\oiv} and {\ov}, we observe that this strong downflow is seen almost co-temporal with the first signatures of the plasmoids. Hence, this strong downflow may be interpreted to be a reconnection outflow counterpart of plasmoids. 

In Fig.~\ref{fig:qs_width_xt}, we find that the line width exhibits structures similar to those observed in the velocity (see Fig.~\ref{fig:qs_velocity_xt}). Here, we use solid pink contours for high intensity (95 percentile of intensity distribution) and dot-dashed white contours for low intensity (10 percentile of intensity distribution). We note that the largest line widths show a strong association with the strongest of flows. This suggests that such strong flows result from \reviewtwo{an ensemble of flows or strong velocity gradient} along the line of sight. We once again observe plasmoid signatures in {\oiv} and {\ov}, while very clear signatures are also seen in {\siiv}.

We note that the plasmoid signatures are observed in the line width maps, but not in the intensity of {\siiv}, possibly due to the LOS effects. {\siiv} emission is dominated by low-lying loops and the bulk of the rising loop. Hence, intensity signatures are dominated by emission from the bulk of the system and not the edges where the plasmoids are formed. However, while the motion of the bulk is rather ordered, the plasmoids exhibit more flow components, which manifests in the width of the spectral lines. Near x = $56$ Mm and $65$ Mm, for example in {\oiv}, we find signatures of excess line widths co-spatio-temporal with excess downflows. These downflows are co-spatial with the foot points of the rising loop (see. Fig.~\ref{fig:qs_lines}). For a temperature of $10^5$ K, $\gamma = 5/3$, we obtain a sound speed of $\approx30$ km/s, {which is lower than the velocities observed in {\oiv} (for example near t = $2600$ s and $2950$ s). Such findings demonstrate the signatures of shock waves, which is observed as enhanced downflows and line widths.

%===============================
\subsection{{\modch}: Results} ~\label{sec:ch_results}
%===********====================

%-====================================
\subsubsection{Dynamics } ~\label{sec:ch_dynamics}
%-====================================
\begin{figure}[h!]
    \centering
    \includegraphics[width=\textwidth]{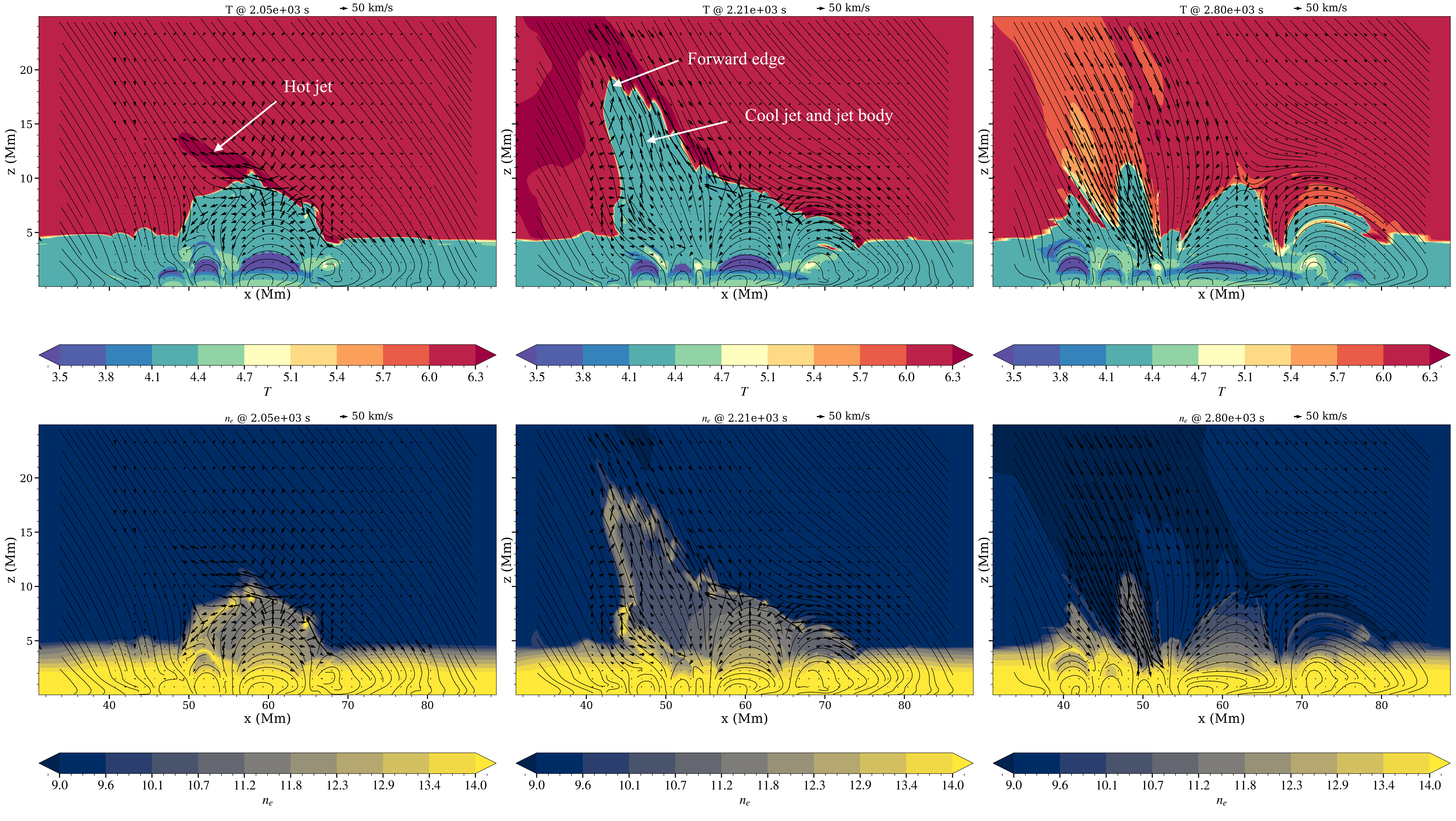}
    \caption{Temperature and density variation for {\modch} at different times, similar to Fig.~\ref{fig:qs_density_temp}. We also refer the reader to animation CH1 of the temperature and density respectively.~\reviewtwo{The animation displays the evolution of the flux sheet from instability to the surge formation, of which the three most important timestamps of interest are displayed in the static image.} }
    \label{fig:ch_density_temp}
\end{figure}
%---------------
We show the density and temperature evolution for the emergence in {\modch} in Fig.~\ref{fig:ch_density_temp}. The emerging flux sheet undergoes reconnection with the ambient magnetic field, resulting in the formation of plasmoids, which are expelled on either side of the reconnection zone. These plasmoids typically have a temperature of $\approx10^4$ K. ~\reviewtwo{The plasmoids} travel outward with a typical speed of $\approx50 - 100$~km/s. These plasmoids collide, as in the {\modqs}, with the ambient atmosphere and result in the formation of a ~\reviewtwo{surge, with a cool body and a hot forward edge}. The \review{resultant} dense, cool jets are at temperatures of $\approx2-5\times10^4$ K, and traveling outward with a speed of $\approx50 - 100$ km/s (for example, between x = 40 and 55 Mm and z = 5 and 25 Mm at $\approx2210$s in Fig.~\ref{fig:ch_density_temp}). The forward edge of the jet corresponding to the region of interaction of plasmoids with the ambient atmosphere has densities of $\approx10^{11}-10^{13}$ cm$^{-3}$, while the body of the jet has densities of $\approx10^{10}$ cm$^{-3}$, as seen in Fig.~\ref{fig:ch_density_temp}. The dense jets are again similar to the slingshot-like motion of plasma from \cite{yokoyama_1996_jetsimulations}. We also observe a hot jet on top of the cool jet, similar to {\modqs} (see Fig.~\ref{fig:ch_density_temp} at $2210$ s).~\review{This jet is seen to be a result of the interaction of the ejected plasmoids with the ambient atmosphere}. The hot jet does not experience \reviewtwo{strong cooling due to its low density}. The hot jet has a characteristic temperature of the order of $\approx10^6$ K. These jets, owing to the background magnetic topology, reach $\approx25$ Mm in height, larger than those seen in {\modqs}. It is also rather clearly seen that at later times ($\approx2.45\times10^3$s), the bulk of the ejected \review{cool} material falls back down. Interestingly, at even later times, the down-flowing material ``splashes'' against the lower-lying plasma, resulting in more ejections along the open flux. This phenomenon may be seen to start clearly at $\approx2.58\times10^3$s and persist till the end of the simulation. This ejection is of lower density ($\approx10^{8}$ cm$^{-3}$) and high temperature ($\approx10^6$ K), while also having very modest velocities ($\approx10$ km/s).

%------------
\subsubsection{Synthetic observables} ~\label{sec:ch_spectrum} 
%------------
\begin{figure}[h!]
    \centering
    \includegraphics[width=\textwidth]{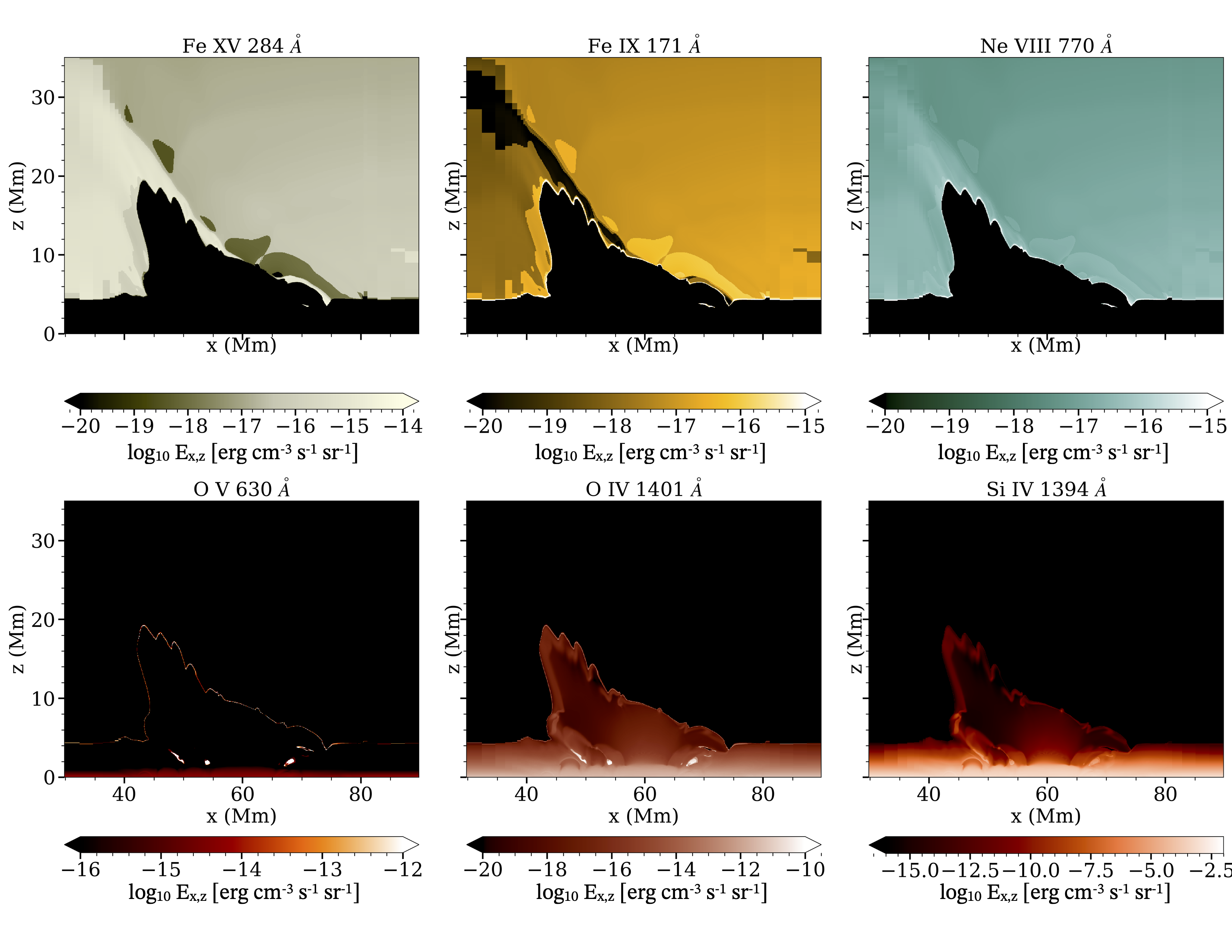}
    \caption{Intensity in different spectral lines for {\modch}, similar to Fig.~\ref{fig:qs_lines}, computed at  t = 2210 s. Notice the distinct gradation in the features of the jet observed in {\siiv} and {\oiv}, and the other hotter lines. We also refer the reader animation CH2~\reviewtwo{where the spectral response corresponding to CH1 is displayed. The static image displays the most important dynamics of interest.} }
    \label{fig:ch_lines}
\end{figure}
%-----------------------------

We consider the snapshot near t = 2210 s and perform spectral synthesis. This is displayed in Fig.~\ref{fig:ch_lines}. The first and most salient observation is the general lack of emission in the hotter lines here when compared to {\modqs}. This is due to low number density, especially at higher temperatures. The hotter lines {\fexv}, {\feix}, and {\neviii} probe the hot jet seen in Fig.~\ref{fig:ch_density_temp}, while also probing the loop top of the newly formed hot loop. \review{We note that the relative enhancement of intensity within the hot jet with respect to the background atmosphere is not extremely large.} {\ov} probes a rather specific region corresponding to the boundary of the hot jet, while {\oiv} also shows signatures of the body of the cool jet and low-lying loops. Bright regions with enhanced emission are seen in {\oiv} and {\ov}, e.g., near x $\approx70$ Mm. These bright regions are formed due to a part of the plasmoid moving along the loop toward the footpoints. The {\siiv} line is sensitive to the \reviewtwo{transition region at the edge} of the cool surge, and structures formed during the bidirectional reconnection flow. The body of the hot loop has cool plasma with some structure, probed by {\siiv}. However, we also note that the background emission coming from low-lying structures is at least $\approx5$ orders of magnitude stronger than the surge signature. 

%------------
\subsubsection{Space-time plots} ~\label{sec:ch_spacetime}
%------------
We now study the evolution of intensity, velocity, and line width of different lines for {\modch} by integrating the spectrum along the vertical ($z$) axis. The evolution is shown from a couple of snapshots just before the onset of reconnection to the end of the simulation. 

\begin{figure}[!h]
    \centering
    \includegraphics[width=\textwidth]{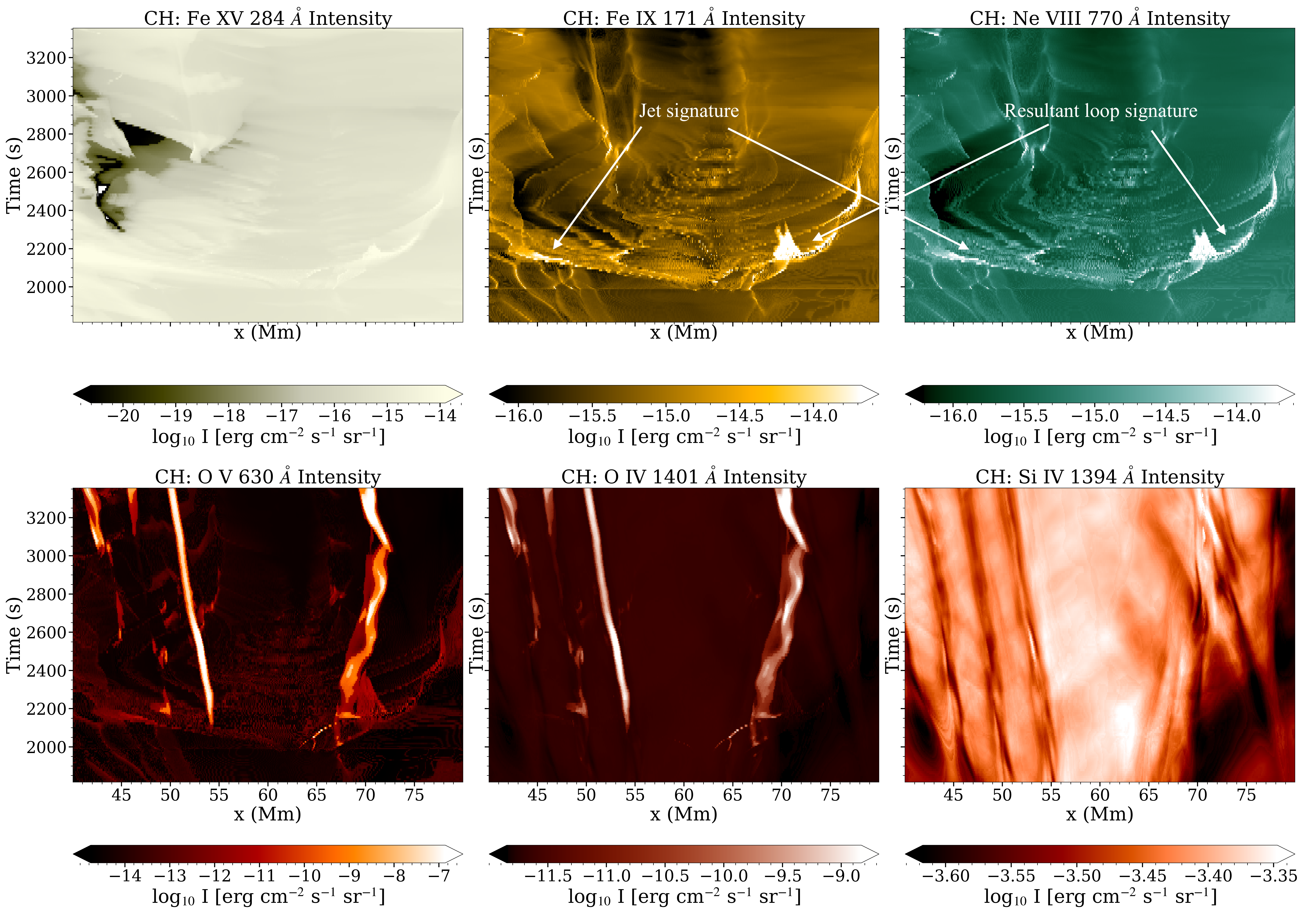}
    \caption{Intensity space-time rasters similar to Fig.~\ref{fig:qs_netflux_xt} but for {\modch}. Note the signatures of the plasmoid shock front in {\neviii} and {\feix}, and minor signatures as small, moving bright dots in {\oiv}, and {\ov}. }
    \label{fig:ch_netflux_xt}
\end{figure}
\begin{figure}[!h]
    \centering
    \includegraphics[width=\textwidth]{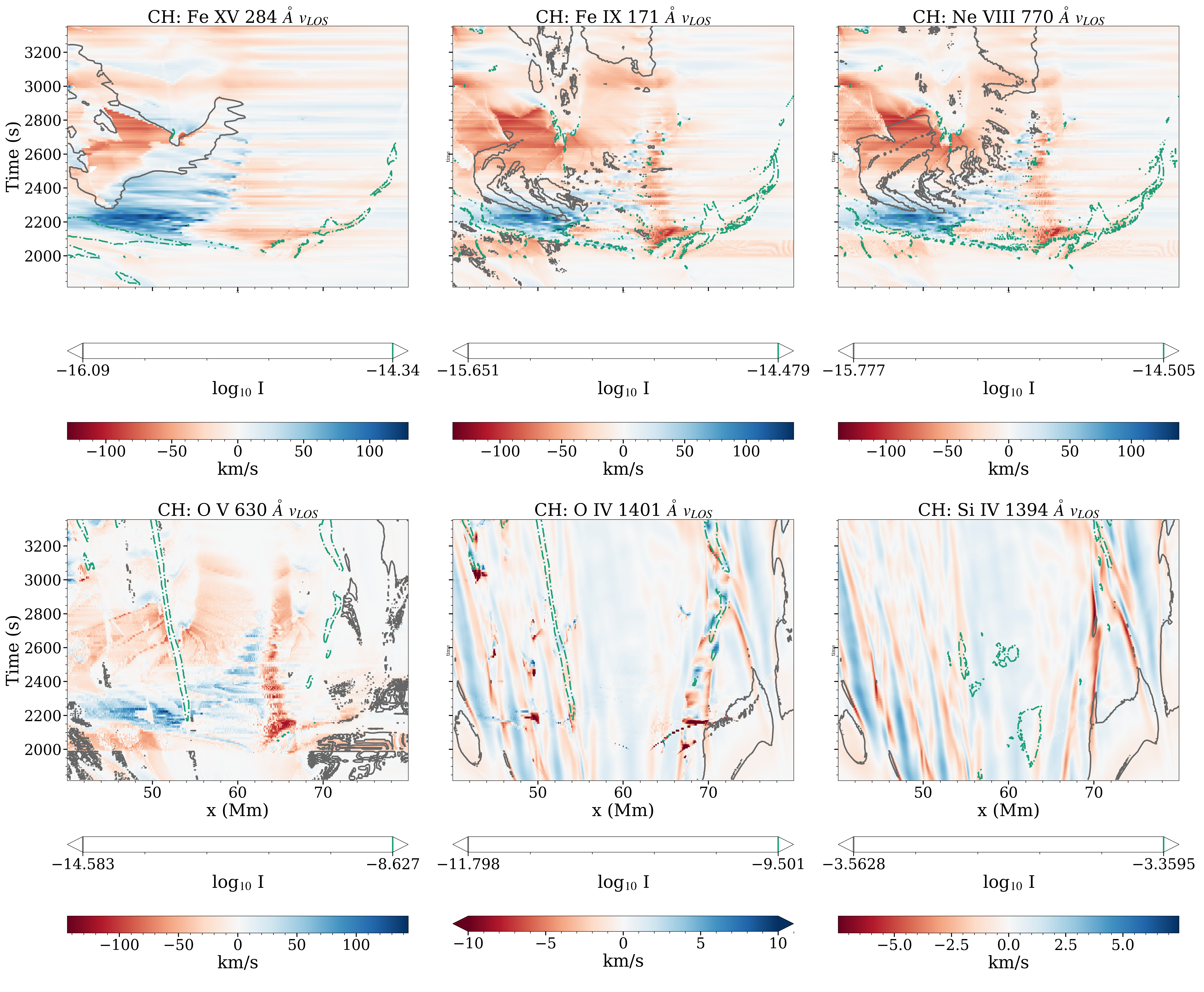}
    \caption{Velocity space-time plot for {\modch} similar to Fig.~\ref{fig:qs_velocity_xt}. Note the similarities in velocity structure of {\siiv} and {\oiv}, and also with {\modqs} {\siiv}. Note also the strong upflow signatures in {\ov}, {\neviii}, {\feix}, and {\fexv}  near t = $2200$ s arising due to reconnection. }
    \label{fig:ch_velocity_xt}
\end{figure}
\begin{figure}[!h]
    \centering
    \includegraphics[width=\textwidth]{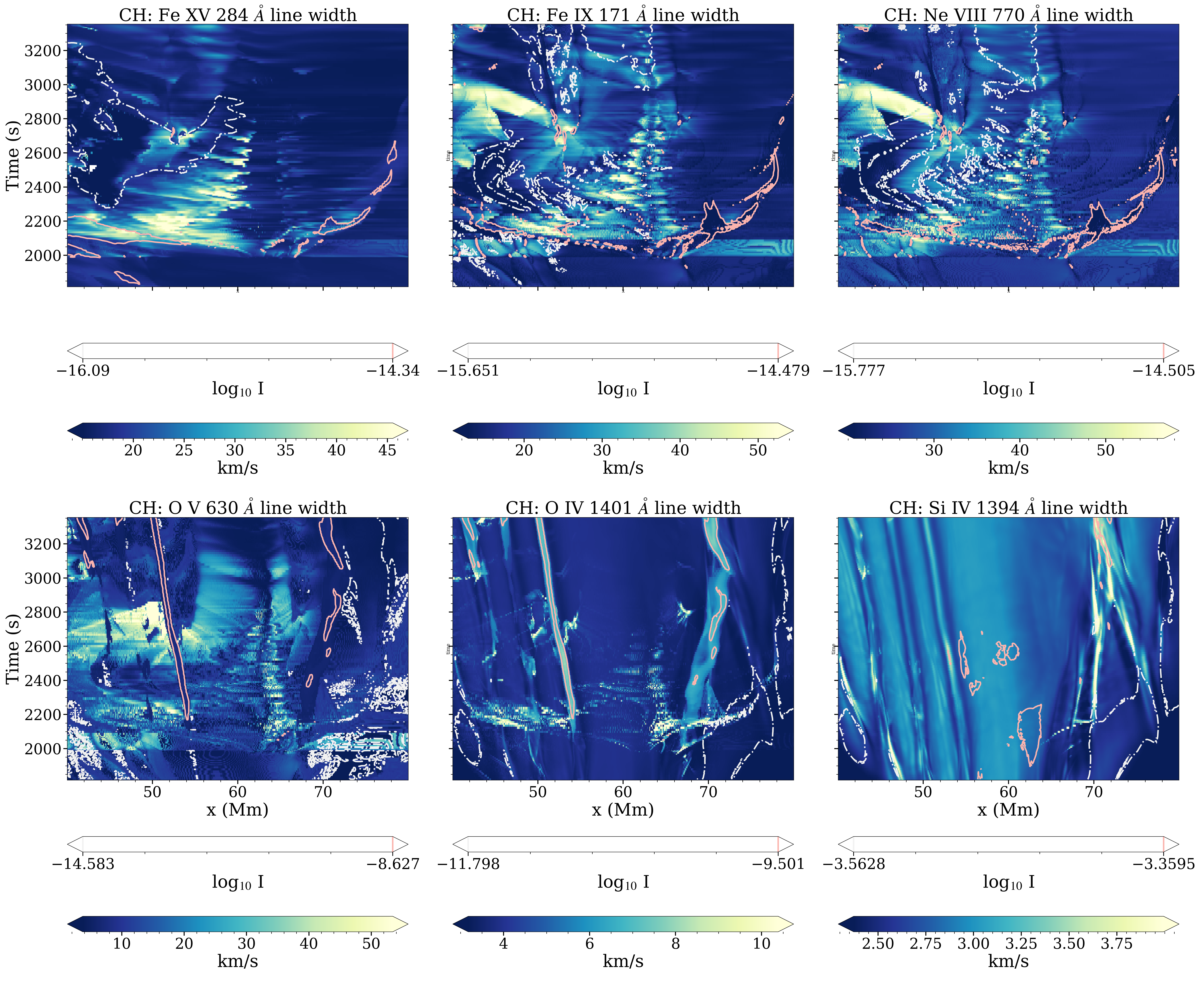}
    \caption{Line width space-time plot for {\modch} similar  to Fig.~\ref{fig:qs_width_xt}. Note the increase in line width magnitude from {\siiv} to {\fexv}. Note also the gradation of structures across lines, similar to Fig.~\ref{fig:ch_velocity_xt}. Finally, note the correspondence between the plasmoid and jet signatures with those in  Fig.~\ref{fig:ch_velocity_xt}. }
    \label{fig:ch_width_xt}
\end{figure}

Fig.~\ref{fig:ch_netflux_xt} demonstrates the evolution of {\modch} features as seen from the top of the box. Similar to {\modqs}, we find that the {\siiv} emission is the strongest amongst all the lines and \reviewtwo{and does not show any strong plasmoid signatures} when compared to {\modqs}. As discussed in \S~\ref{sec:ch_spectrum}, this occurs due to the dominance of ``low-lying loop'' emission arising from the emergent flux sheet. As we noted in \S~\ref{sec:ch_spectrum}, we do not observe any strong signatures of the surge along the line of sight, unlike those seen in Fig.~\ref{fig:ch_lines}.

In {\oiv} and {\ov} we notice two distinct bright streaks starting from x = $54$ and $68$ Mm and t $\approx2100$ s. Similar streaks are also observed at t $\approx3000$ s near x = $45$ Mm. Such streaks are time-varying signatures of the bright dot structures seen in Fig.~\ref{fig:ch_lines}. These structures, as we noted, form due to the accumulation of some of the reconnected plasma at the lower atmosphere, resulting in enhanced densities. In {\neviii} and {\feix}, we notice a distinct signature of plasma propagation (along with an intensity enhancement) from t $\approx2000$ s~\review{from x $\approx$62 Mm to x$\approx$45 Mm.} These propagating structures are the `hot' boundaries of the cool surge.~\review{On the other hand, we also see running intensity enhancement from x$\approx$62 Mm to $\approx$75 Mm. These enhancements are signatures resulting from the plasmoids moving in the direction opposite to the surge, towards the newly formed hot loop.} Since the background emission from low-lying structures is minimal in {\neviii} and {\siiv} in {\modch}, we find much stronger signatures of the initial phase of the surge. The {\fexv} line does not show any particular signatures in this case, except for a reduction in intensity near t = $2300$ -- $2900$ s, and between x = $43$ and $55$ Mm. \review{This intensity reduction can be seen from the animation CH2, associated with Fig.~\ref{fig:ch_lines} to be just after the eruption of the surge, and through its fall back towards the photosphere. This eruption-and-fall results in a reduction in density of plasma corresponding to the sensitivity of the hot lines, and hence an intensity reduction.} The reduced emission signatures are seen even in {\feix} and {\neviii} lines co-spatially and co-temporally.

Unlike {\modqs}, we do not see explicit signatures of plasmoids here. In {\oiv} and {\ov}, mild signatures are seen as moving small dots from (x,t) = ($63$ Mm, $2000$ s) to ($68$ Mm, $2100$ s). These plasmoids are much smaller and do not grow similar to {\modqs}. However, note that the initial model atmosphere has the transition region at twice the height in {\modch} when compared to {\modqs}. Mild signatures of these plasmoids -- co-spatial and co-temporal -- are also seen in the other lines (including {\fexv}) as propagating small bright dots. The hot emission corresponds to the `high-temperature edge' of the plasmoids, while the plasmoids themselves are cool due to high density.

In Fig.~\ref{fig:ch_velocity_xt}, we show the vertical velocity, with the same scheme as in \S~\ref{sec:qs_spacetime}, with plasma moving towards the observer in blue color and moving away in red. We also overlay intensity contours, black solid standing for low (10 percentile of intensity distribution) and green dot-dashed for high intensities (95 percentile for intensity distribution). We observe some rather interesting structures in these velocities. In {\siiv}, we find alternating blue and redshifted patterns, similar to {\modqs}. Some of the intensity deficit regions show an association with the downflows away from the observer - for example, x = $69$ Mm between t = $2700$ -- $3000$ s. The bright regions, on the other hand, correspond to the surge and underlying short loops as seen in Fig.~\ref{fig:ch_lines}. {\oiv} shows a significant contribution of both the general loop dynamics and clear signatures of plasmoids (near x = $65$ Mm, t = $2000$ s). These plasmoids are mostly moving away from the observer when formed towards the right of x = $62$ Mm. The plasmoids forming in this region would be guided along the lower end of the rising loop and the background field, towards the bottom of the domain. However, the plasmoids formed near (x = $60$ Mm, t = $2000$ s) Mm move upwards and are seen in blueshifts. These signatures are not very strong in {\oiv}, but the emission from the hot edge of the plasmoids is seen clearly in all the other spectral lines. ~\review{We note that the cool jet does not show prominent strong signatures in {\oiv} and {\siiv} owing to the much smaller emission from the jet with respect to the low-lying loops, as also evident from Fig.~\ref{fig:ch_lines}.}

The hotter lines all show a prominent upflow from (x,t) = ($60$ Mm, $2000$ s) to ($45$ Mm, $2150$ s) due to the jetting activity, also resulting in a strong upflow near x = $50$ Mm, t = $2200$s and lasting for $\approx150$~s. At later times (after t  = $2600$ s), the system relaxes, resulting in downflows in hotter lines. The prominent upflow lasting $150$~s is seen just after the brightest structures are seen in these lines (note that the y-axis is time). However, the downflow seen in hotter lines corresponds to the intensity deficit structures. The intensity deficit structures, as we have seen in Fig.~\ref{fig:ch_netflux_xt}, arise due to ``clearing-up'' of hot plasma by the cool plasma, resulting in reduced plasma where the hot lines are sensitive. Similarly, the plasmoids moving down, seen as downflows near x = $69$ Mm \reviewtwo{(near t = $2000$ - $2150$ s)}, result in localized strong downflows too. The downflow signatures are seen just after the brightest structures are seen~\review{moving from x$\approx$62 Mm to $\approx$75 Mm.}

In Fig.~\ref{fig:ch_width_xt}, we showcase the line width overlaid with intensity contours (pink solid at 95 percentile intensity and white dot-dashed for low-intensity structures at 10 percentile). We find that the line width exhibits structures similar to the velocity in Fig.~\ref{fig:ch_velocity_xt}. Many of the largest line widths show a strong association with the strongest of flows, similar to {\modqs}. Interestingly, we find the line width enhancements to be co-spatial and co-temporal to downflowing plasma, for {\siiv} and {\oiv}. The cool plasma directed away from the observer undergoes collision with the underlying plasma, resulting in multiple flow components along the line of sight. In the hotter lines, the surge exhibits strong upflows, resulting in co-spatial and co-temporal enhancements in line widths. We also observe signatures of plasmoids in {\siiv} and {\oiv} corresponding to the downflowing plasma, while the upflowing plasmoids show signatures only in {\oiv}. We also find a strong association between the high-intensity structures in {\oiv} and strong line widths. Since the high intensities correspond to the accumulated plasma, the collision of downflowing plasma results in emission from multiple flow structures, manifesting as enhanced line widths.

%=======================================
\subsection{{\modqs} and {\modch}: Comparison during the surge~\label{sec:upflow}}
%=======================================
We have studied the dynamics in different spectral lines as space-time plots in \S\ref{sec:qs_spacetime} and \ref{sec:ch_spacetime}. We have seen differences in intensity, velocity, and line width signatures of flux emergence in CH and QS.~\review{We now seek to understand the association between the synthesized spectral properties themselves, and with the magnetic field ({\bmag}$\coloneqq|B_z(z=0)|$),~\reviewthree{qualitatively mimicking analysis between such observables}. Performing such a study also} provides statistical, quantitative differentiation between the two regions. Hence, we next study the specific dependence of these on {\bmag} and relations between the velocities derived from different lines. We first consider the full domain at timestamp \review{t $\approx2200$s in {\modch} and t $\approx2300$s in {\modqs}, to capture similar stages of evolution} in both the models. \review{This timestamp corresponds approximately to the strong upflows in {\modch}, and a similar stage of jet in {\modqs}}. We study the dynamics during the return flow in {\modch} in \S~\ref{sec:downflow}. % \. The dynamics of the surge in {\modch}, for instance, is very complex. However, the observables one generally obtains are the spectra, and the photospheric magnetic field. Typical observations, especially in CH and QS, do not provide magnetic field inferences at any other heights in the atmosphere. 

~\reviewtwo{We jointly sort these moments as a function of {\bmag}, and} bin the moments in deciles of {\bmag} -- i.e., every 10\% of ~\reviewtwo{the sorted} {\bmag} values are considered in one bin. We report the mean value of the moment in that {\bmag}, while the error bars correspond to the standard error on the mean. The standard error is defined as $\sigma/\sqrt{N}$, where $\sigma$ is the standard deviation for the samples present in the bin, and $N$ is the total number of samples. The standard error captures the uncertainty in the estimation of the mean, and we are only interested in the average behavior of these line moments with the mean. The scatter plot is shown in Fig.~\ref{fig:qs_ch_modb}. We also display the Pearson correlation between the binned {\bmag} and mean values of the moments per {\bmag} bin as $\rho_{CH}$ for CH and $\rho_{QS}$ for QS in the inset. \reviewfour{We emphasize that our science of interest is only in an average sense since our evaluation of CH and QS in this work is to enable a qualitative comparison with {\ppsi} and {\ppmg}. Quantifying the association between the averages would hence necessitate correlation to be evaluated on the binned quantities. Furthermore, we emphasize that the general dynamics of the system is very complex, especially in regions with low {\bmag}, as  evidenced from Fig.~\ref{fig:qs_density_temp} and ~\ref{fig:ch_density_temp}. Since we are primarily interested in the dynamics of the jet, we consider the correlations only between the binned quantities \reviewfive{to minimize the impact of potentially local dynamics in the weak field regions from dominating the statistics}. } 

We consider a correlation to be strong if it is $\geq0.8$ with a p-value of $\leq10^{-3}$, with a similar definition for an anti-correlation. We define the correlation to be mild if the values are $<0.8$ but $>0.1$, and negligible otherwise. Since we consider these correlations as either mild or negligible, we do not check for the associated p-value. The p-value is a measure of the probability of accepting the Null hypothesis, which in this case is that the two parameters are uncorrelated. Hence, a smaller p-value implies a higher statistical significance of the correlation. We, ~\reviewtwo{however, emphasize} that the interpretation of the results would be an interplay of the correlation values and associated standard errors computed per bin, ~\reviewfour{for any study of variations of a highly-uncertain mean is rather meaningless}.

\begin{figure}[ht!]
    \centering
    \includegraphics[width=\textwidth]{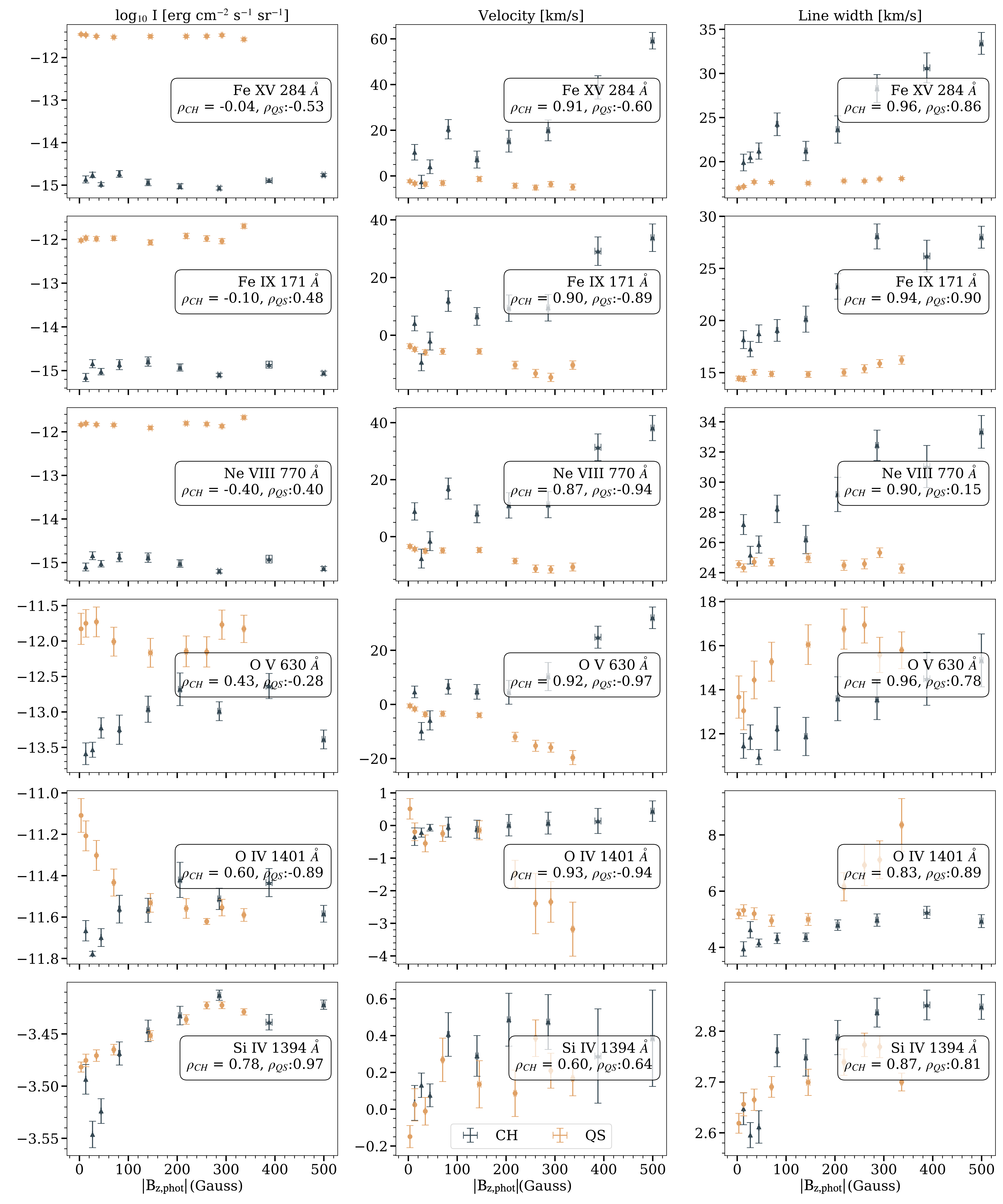}
    \caption{Intensity ($I$; left column), velocity (central column, downflow is negative), and line widths (right column) as a function of underlying photospheric magnetic field ({\bmag} at z=0), binned for every 10\% of {\bmag}. Black color corresponds to {\modch}, and orange to {\modqs}. ~\review{We display the Pearson correlation between the binned {\bmag} and the mean moment per bin for {\modqs} (as $\rho_{QS}$) and {\modch} (as $\rho_{CH}$), in the inset for each panel}. Computed at t$\approx2200$ s for {\modch} and at $\approx2300$ s for {\modqs} across the whole domain.}
    \label{fig:qs_ch_modb}
\end{figure}
%-----------------------------
%------------------------------
\begin{figure}[h!]
    \centering
    \includegraphics[width=\textwidth]{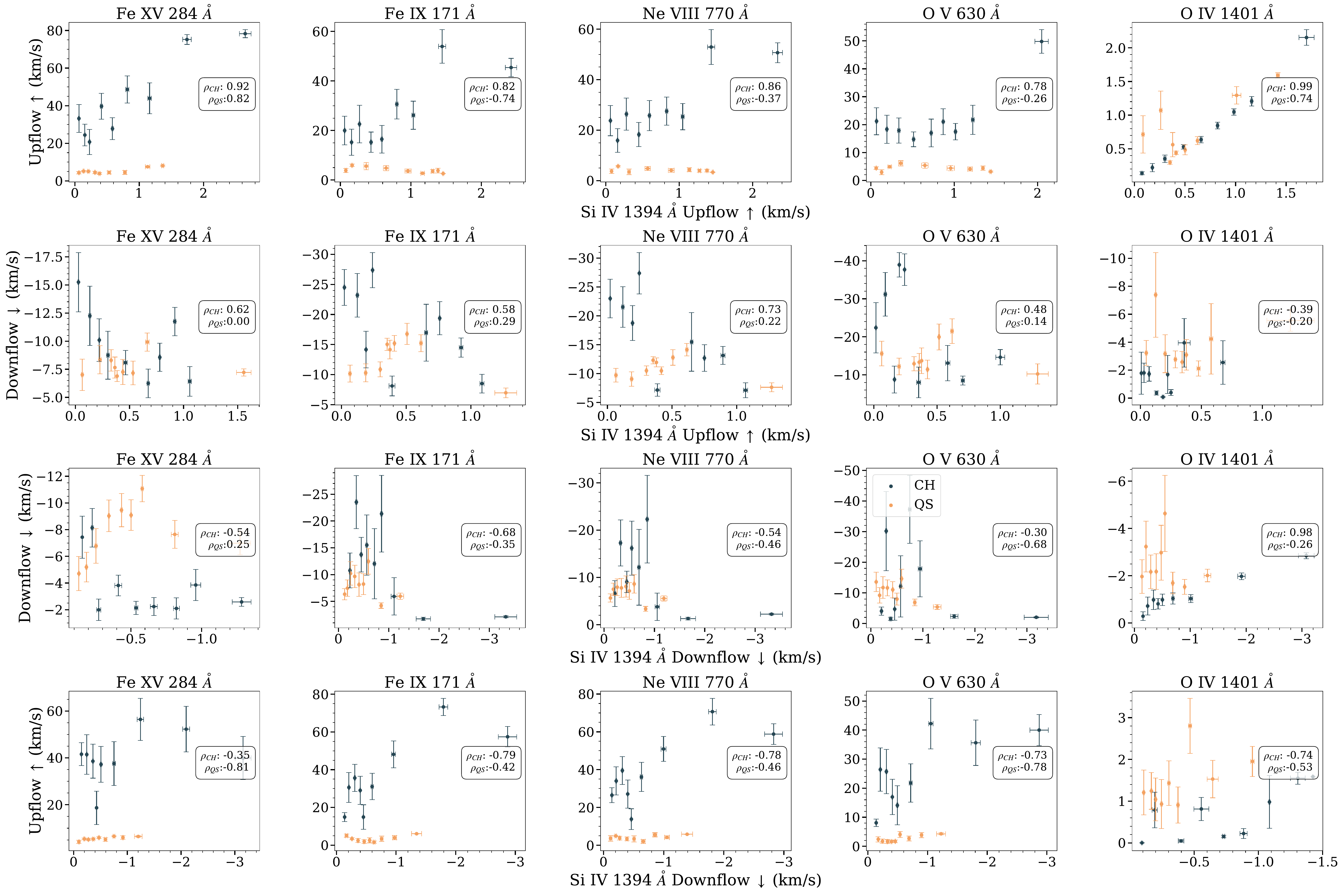}
    \caption{Scatter plots between co-spatial velocities in {\siiv} and {\fexv}, {\feix}, {\neviii},{\ov}, and {\oiv} lines. The first two rows correspond to points with upflows in {\siiv}, while the last two correspond to downflows. Each column corresponds to a particular line, and the sense of velocity for each of the columns is depicted on the left column. Black color corresponds to {\modch}, and orange to {\modqs}. ~\review{We display the Pearson correlation between the binned {\siiv} velocity and the mean velocity per bin of the other lines  for {\modqs} (as $\rho_{QS}$) and {\modch} (as $\rho_{CH}$), in the inset for each panel}. Computed at t$\approx2200$ s for {\modch} and at $\approx2300$ s for {\modqs} across the whole domain. }
    \label{fig:qs_ch_intercorrelations}
\end{figure}
%-----------------------------

Fig.~\ref{fig:qs_ch_modb} reveals several features regarding the differences in the dynamics of {\modqs} and {\modch}. First, the intensities in QS are larger than those in CH, as expected from our model atmosphere. \review{In {\modqs}, we find the {\siiv} intensities to have a strong correlation with {\bmag}, which changes to strong anti-correlation in ~\reviewtwo{{\oiv} and mild correlations in the other lines}. In {\modch}, however, {\siiv}, {\oiv}, and {\ov} show mild correlation between intensity and {\bmag}, with the correlation getting weaker with formation temperature. {\neviii} shows a mild anti-correlation between intensity and {\bmag}, which becomes almost negligible in hotter lines.} Furthermore, we note that the intensities are most consistent between {\modqs} and {\modch} in {\siiv}, while the consistency reduces in other lines. 

The velocities in Fig.~\ref{fig:qs_ch_modb} reveal an interesting picture. In {\siiv}, both {\modqs} and {\modch} show modest upflows of $\leq1$ km/s,~\review{ and show a mild correlation with {\bmag}}. In {\oiv}, we find modest downflows in {\modqs}, and \review{almost consistent with 0 km/s. Note that while the correlation depicts a rather large value ($\rho_{CH}=0.93$), the errorbars result in the velocities being almost consistent with 0}. However, in hotter lines, we find that {\modqs} predominantly shows downflows (note again that a negative velocity is towards the photosphere in these simulations),~\review{and these velocities are strongly correlated with {\bmag}, ~\reviewthree{except a mild correlation in {\fexv}}}. In {\modch}, we find the opposite, i.e., the surge is seen as an enhanced upflow from the region~\review{as expected from Fig.~\ref{fig:ch_velocity_xt}}.~\review{The velocities of these jets are strongly correlated with {\bmag}.} The magnitude of these velocities also increase from {\siiv} (less than 1 km/s) to {\ov} (20 km/s at max) and hotter lines ($\approx40$ km/s at max).

The line widths in  Fig.~\ref{fig:qs_ch_modb} show an increase with both the approximate line formation temperature, and the underlying {\bmag}. In {\siiv}, these widths are consistent till {\bmag} = 100 Gauss, and then diverge. Except {\oiv} and {\ov}, we find that the line widths are larger in {\modch} when compared to {\modqs} in all the other lines,~\review{while also showing strong correlation ($>0.8$) with {\bmag}. We also note that the correlation is weaker in {\modqs}.}.  This is interesting because the velocity and line width structures in {\modqs} (as seen in Fig.~\ref{fig:qs_velocity_xt} and Fig.~\ref{fig:qs_width_xt}) in {\oiv} are closer to the hotter lines ({\ov}, {\neviii}). In {\modch}, these structures (Fig.~\ref{fig:ch_velocity_xt}, and Fig.~\ref{fig:ch_width_xt}) are closer to the structure in {\siiv}. Thus, the same spectral line seems to probe different regions in different topologies, ~\reviewthree{which results as} the differences in the line widths. 

We now study the relation between the velocities inferred from different spectral lines, which is a measure of coupling across different parts of the solar atmosphere. For this purpose, we consider the velocity in different lines~\review{integrated along $z$}, but at the same spatial location~\review{(i.e. at the same $x$)}. We consider the velocities in ~\review{multiple sets, such as (upflows in one line, upflows in another line), (upflows in one line, downflows in another line), and so on for all four combinations. We consider `reference line' in this set as the} {\siiv} ~\review{line, for it} forms at the lowest temperature in our setup. We consider points that show upflows in both {\siiv} and other lines, downflows in both {\siiv} and other lines, and upflows/downflows and downflows/upflows in {\siiv} and other lines respectively. These combinations are to study the \review{(i).}persistence of upflows, \review{(ii).} persistence of downflows, and \review{(iii).} probe the presence of bidirectional flows between the transition region and the corona,~\reviewfour{by considering spectral lines \reviewfive{with different approximate formation temperature}}. These combinations are then binned in deciles (every 10\%) of {\siiv} velocity to bring out the underlying average variation. The uncertainties depict the standard error over the mean, for we are only interested in the mean behavior in this work. This analysis is presented in Fig.~\ref{fig:qs_ch_intercorrelations}, where the black color is for {\modch} and orange for {\modqs}.~\review{We have once again computed the Pearson correlation between the binned {\siiv} velocity and the mean velocity per bin of the other lines, and displayed them in the inset for each panel.}

We find very interesting relations between these flows. ~\reviewtwo{From the first row of Fig.~\ref{fig:qs_ch_intercorrelations}, we} first find that the upflows in {\siiv}~\review{in {\modch}} are well correlated with upflows in the other lines. This is evidenced by the strong correlation values ($>0.8$,~\reviewtwo{except for {\ov} which has a correlation just short of 0.8}) between the upflows. ~\reviewtwo{In {\modqs}, only the {\oiv} and {\fexv} upflows show mild and strong correlations with upflows of {\siiv}. We further note that the upflows in {\fexv} are very modest, at $\approx5$ km/s. All other lines show varying levels of anti-correlation. Considering the velocity values and the associated errorbars, we find that {\ov}, {\neviii}, and {\feix} do not show any significant association between the upflows.} In general, the upflows are larger in {\modch} over {\modqs}. However, in {\oiv}, we see that {\modqs} shows larger upflows for a given speed in {\siiv}. This arises from {\oiv} probing much higher layers of the atmosphere in {\modqs} over {\modch}. This is seen from the similarities in the velocity structure between {\oiv} and {\ov} in {\modqs} but between {\oiv} and {\siiv} in {\modch}  in Fig.~\ref{fig:qs_velocity_xt} and Fig.~\ref{fig:ch_velocity_xt}, respectively. The upflows signatures are however absent in {\modqs} in {\fexv}. 

Next, the downflows~\review{in different lines do not show a very strong correlation with {\siiv} upflows,~\reviewtwo{in the second row of Fig.~\ref{fig:qs_ch_intercorrelations}}. The strongest anti-correlation is seen between the {\neviii} downflow and {\siiv} upflows ($\approx0.73$) for {\modch}. Despite a similar anti-correlation between the {\oiv} velocities in {\modqs}, we do not consider them to be significant due to their very large errorbars. }

Third, we find very mildly correlated~\review{and even anti-correlated} downflows between the hotter lines and {\siiv}~\reviewtwo{in the third row of Fig.~\ref{fig:qs_ch_intercorrelations}}. The downflows are stronger in the hotter lines in {\modqs} than {\modch}. Finally,~\reviewtwo{from the fourth row of Fig.~\ref{fig:qs_ch_intercorrelations}}, we find that the upflows in all lines show a ~\reviewtwo{mild to strong} correlation with downflows in {\siiv}. The upflows are once again stronger in {\modch} over {\modqs}. We note that correlation values are negative to indicate the difference in signs of the two velocities in consideration. We further find the strongest correlation between the upflows in {\feix} and {\neviii}, and downflows in {\siiv} in {\modch}. ~\reviewtwo{We find that while {\modqs} does not show strong correlations in general, {\fexv} upflows show a strong correlation with {\siiv} downflows, though we note that the velocities barely change $\approx5$ km/s in {\fexv}.} In summation, these results are strongly reminiscent of correlations obtained lower in the atmosphere by \cite{Upendran_2022_CHQS}. 

%=======================================
\subsection{{\modqs} and {\modch}: Comparison during return flow ~\label{sec:downflow}}
%=======================================
%--------------------------
\begin{figure}[h!]
    \centering
    \includegraphics[width=\textwidth]{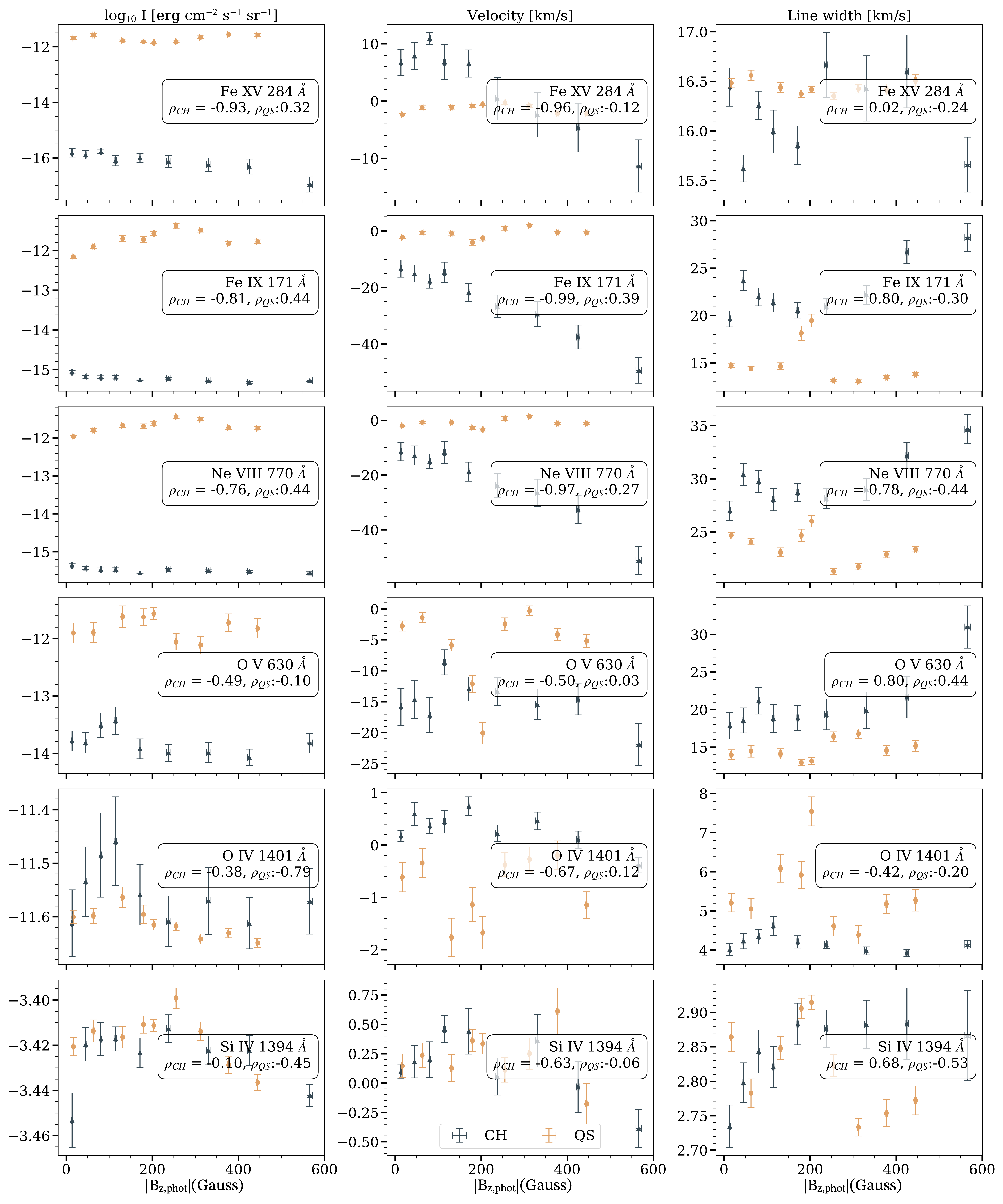}
    \caption{Intensity ($I$; left column), velocity (central column, downflow is negative), and line widths (right column) as a function of underlying photospheric magnetic field ({\bmag} at z=0), binned for every 10\% of {\bmag}. Black color corresponds to {\modch}, and orange to {\modqs}.~\review{We display the Pearson correlation between the binned {\bmag} and the mean moment per bin for {\modqs} (as $\rho_{QS}$) and {\modch} (as $\rho_{CH}$), in the inset for each panel}. Computed at t $\approx2700$ s.}
    \label{fig:qs_ch_modb_down}
\end{figure}
%-----------------------------
\begin{figure}[h!]
    \centering
    \includegraphics[width=\textwidth]{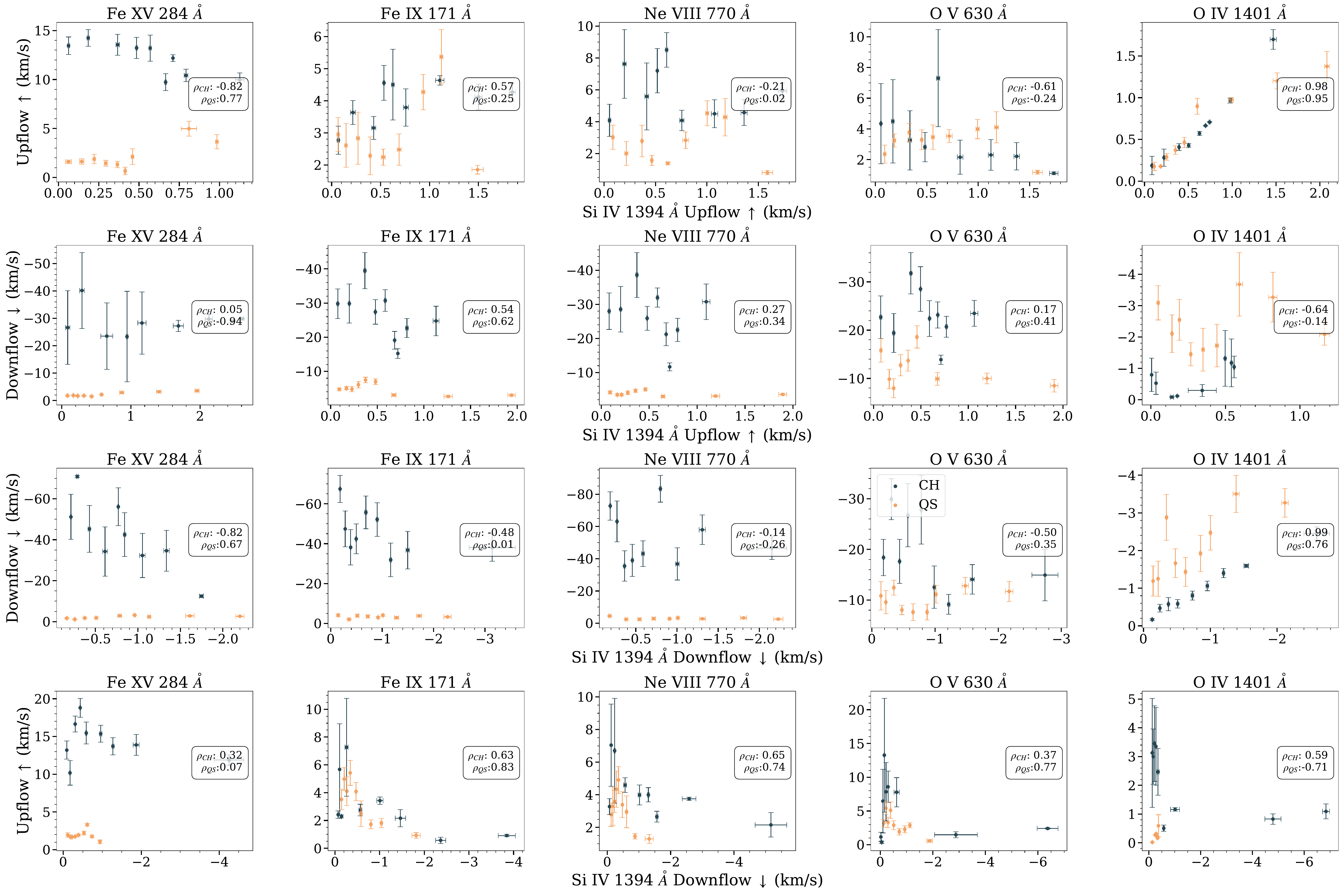}
    \caption{Scatter plots between co-spatial velocities in {\siiv} and {\fexv}, {\feix}, {\neviii},{\ov}, and {\oiv} lines. The first two rows correspond to points with upflows in {\siiv}, while the last two correspond to downflows. Each column corresponds to a particular line, and the sense of velocity for each of the columns is depicted on the left column. Black color corresponds to {\modch}, and orange to {\modqs}.~\review{We display the Pearson correlation between the binned {\siiv} velocity and the mean velocity per bin of the other lines  for {\modqs} (as $\rho_{QS}$) and {\modch} (as $\rho_{CH}$), in the inset for each panel}. Computed at t$\approx2700$ s.}
    \label{fig:qs_ch_intercorrelations_down}
\end{figure}
%-----------------------------
We now consider the differences between the two models at t $\approx2700$~s, when the persistent downflows are seen in {\modch}.~\review{We once again consider the full box for this analysis.} During this time, we have seen co-spatial strong downflows and reduced intensity from Fig.~\ref{fig:ch_velocity_xt}. Considering the moments as a function of {\bmag} in Fig.~\ref{fig:qs_ch_modb_down}, we find that the intensities in {\modqs} are once again larger than those in {\modch}, except in {\oiv} where the ~\reviewthree{intensity differences appear inconclusive within the standard errors}. We find the {\modch} intensities to be anti-correlated with {\bmag}, with the anti-correlation increasing with approximate line formation temperature~\reviewtwo{, where the anti-correlation changes from mild in {\siiv} to strong in {\fexv}.}. We also find either a mild correlation, or a mild anti-correlation for {\modqs} intensities with {\bmag}.

We find the flow velocities to generally increase~\reviewtwo{in absolute value} with the approximate formation temperature of the line, except for {\fexv}. We find strong downflows in all the hot lines from {\modch}. These downflows also show a strong correlation with {\bmag}. In {\modqs}, we do not see such strong flows correlated with {\bmag}. In {\oiv}, we yet again find {\modch} to have velocities $\approx0$, while {\modqs} exhibits some uncorrelated downflows. In {\siiv}, we see that the flow changes sign from upflows at low {\bmag} to downflows at large {\bmag}, while exhibiting a correlation in {\modch}.

The line widths in all of the spectral lines exhibit either a~\reviewtwo{mild} anti-correlation, or ~\reviewtwo{a mild correlation in {\ov}} with {\bmag} in {\modqs}. In {\modch}, the line widths exhibit a positive correlation with {\bmag} for {\ov}, {\neviii} ~\reviewthree{(albeit at 0.78 correlation)}, and {\feix}. We note that while the {\siiv} line width shows a \reviewthree{mild} correlation with {\bmag}, the errorbars are far too large to make any substantial observation.

\review{In general, we note that }the plasma dynamics is {\modqs} has reached a near steady-state, and does not have specific dependence on the underlying photospheric magnetic field. In {\modch}, the return flow is purely the strong upflow plasma falling back towards the photosphere. %Hence, we see dependence of the line moments on {\bmag} in {\modch}.

We next also consider the correlation between velocities of different spectral lines during this phase in Fig.~\ref{fig:qs_ch_intercorrelations_down}. First, ~\reviewtwo{from the first and third rows,} we find a strong correlation between upflows (downflows) in {\siiv} with upflows (downflows) of {\oiv}~\review{for both {\modch} and {\modqs}}. This strong correlation~\review{occurs} due to {\siiv} and {\oiv} probing very similar parts of the atmosphere.~\reviewtwo{From the first row of Fig.~\ref{fig:qs_ch_intercorrelations_down}, we find that the other lines either show a negligible or a mild correlation in {\modqs}. Note that the high correlation in {\fexv} may be misleading -- the relationship between the velocities, considering the standard errors, is not linear. From the second row, we find that the downflows in hot lines do not have any strong correlation with {\siiv} upflows in {\modch}. In {\modqs}, we find strong anti-correlation only between {\fexv} downflow and {\siiv} upflow. We note once again that while the correlation is strong (note the sign due to the sign of the velocities), the velocities in {\fexv} are very small. }

~\reviewtwo{From the third row of Fig.~\ref{fig:qs_ch_intercorrelations_down}, we find {\modqs} to exhibit mild to negligible correlations across all lines. The most significant correlations occur between the downflows in {\oiv} and {\siiv}, noting that the {\fexv} downflows are yet again very small. In {\modch}, we find strong correlation between the downflows in {\oiv} and {\siiv}. The other lines generally show mild anti-correlations, except for the strong anti-correlation in {\fexv}.}

From the final row of Fig.~\ref{fig:qs_ch_intercorrelations_down}, we note that {\modqs} shows a mild correlation between {\oiv} upflows and {\siiv} downflows, while showing~\reviewthree{anti-correlation with other lines (strong in {\feix}, and} almost no correlation in {\fexv}. We however note that the upflows in {\feix} shows a strong positive correlation value (anti-correlation, considering the velocity direction) with downflows in {\siiv} for {\modqs}. Such an anti-correlation, albeit weaker, is seen for the {\neviii} and {\ov} velocities too. In {\modch}, we generally find mild correlations between the flows. Even the strongest of correlations have large errorbars, making it very difficult for interpretation. Moreover, most of the velocity combinations do not show any strong correlations like those seen in Fig.~\ref{fig:qs_ch_intercorrelations}. This points towards major differences in the coupling across the atmosphere during upflows (~\reviewthree{as discussed in} \S.~\ref{sec:upflow}) and downflows discussed here. 

%~\review{We now move onto a summary of our work and interpretation of our findings from this analysis.} %Hence, we suggest that while the return plasma flow retains some dependence on the {\bmag}, the hot plasma, which gives rise to emission above the formation of {\oiv}, is effectively decoupled from the cool plasma probed by {\siiv}. 

%==============================================
\section{Summary and discussion}~\label{sec:discussion}
%==============================================
In this work, we have performed 2.5D MHD simulations of flux emergence. The core idea is to compare the dynamics of flux emergence~\review{in} CH and QS background field setup.~\review{We perform this comparison by synthesizing observables in spectral lines probing the transition region and solar corona. We also perform statistical analysis on associations between these observable parameters.} We find that the local dynamics of reconnection, and the properties of plasmoids are similar in both~\reviewtwo{the CH and QS} cases. The salient results from our analysis are:
\begin{enumerate}
    \item Reconnection results in formation of cool ($\approx10^4$ K) plasmoids, $\approx0.5$ Mm in size, that move out with velocities of $50-100$ km/s in both {\modqs} and {\modch}.
    \item Cool ($\approx2-5\times10^4$K), dense ($\approx10^{11}-10^{13}$ cm$^{-3}$ ) jets are ejected as a result of reconnection. However, hot ($\approx10^6$ K), less dense ($\approx10^8-10^{10}$ cm$^{-3}$) jets are also seen to form at the `coronal' interface of the cool jet, riding on top of the cool jet.
    \item The jets are predominantly horizontal in QS, while they are guided by the slanted field in the CH setup. They leave signatures in intensities, velocities, and line widths, with excess upflows in {\modch}.
    \item ~\review{When we compare the upflow phase of {\modch} with the same timestamp of {\modqs}, we find that} {\modch} shows lower intensities, larger upflows, and greater line widths in ~\reviewtwo{{\ov}, {\neviii}, {\feix}, and {\fexv}} when compared to {\modqs}~\reviewtwo{(see Fig.~\ref{fig:qs_ch_modb})},~\reviewthree{except {\ov} line widths}. ~\reviewtwo{We find the intensities in these lines show mild correlation or anti-correlation with {\bmag} in both {\modqs} and {\modch}. In {\oiv}, we find a strong anti-correlation between intensity and {\bmag} in {\modqs}, while a weak correlation in {\modch}. In {\siiv}, we find a strong correlation between the intensity and {\bmag} in both the regions (see Fig.~\ref{fig:qs_ch_modb}).} We find a strong linear~\reviewthree{association} between velocities, line widths and {\bmag} for {\ov}, {\neviii}, {\feix} and {\fexv} in {\modch},~\reviewtwo{while the line widths in {\siiv} and {\oiv} show a strong correlation with {\bmag}}.~\reviewtwo{We note that the velocity correlation is weak in {\siiv} for {\modch}, and strong for {\oiv} but consistent with 0.} In {\modqs}, we however find downflows strongly correlated with {\bmag}~\reviewthree{for {\oiv}, and a mild correlation for {\siiv}}. However, these differences between {\modqs} and {\modch} almost vanish in the {\siiv} line. 
    
    \item We study the correlation between {\siiv} velocities and those from the other lines~\reviewtwo{in Fig.~\ref{fig:qs_ch_intercorrelations}. During the upflow phase,} we find a strong correlation between the~\reviewtwo{upflows in hotter lines and {\siiv} upflows, in {\modch}. Such a correlation is not consistently observed in {\modqs}. In {\modch}, we do not find general strong correlations except for the upflows in {\neviii} and {\feix} and downflows in {\siiv}, and the downflows in {\oiv} and {\siiv}. In general we also find the correlation values in {\modqs} are small, except for the upflows in {\fexv}, where the velocity values are $\approx5$ km/s.}
    \item ~\reviewtwo{In Fig.~\ref{fig:qs_ch_modb_down}, we compare the  line moments from {\modqs} and {\modch} with {\bmag} during the downflow phase of {\modch}. In general, we we obtain excess intensities in {\modqs} except in {\oiv}, while the correlation values are mild. We obtain stronger downflows in {\modch}, with the correlation value getting approximately stronger from ~\reviewthree{{\ov} to {\feix}}. Such a trend is not seen in {\modqs}. We obtain a strong correlation with line widths only in {\modch} for the {\ov}, {\neviii}, and {\feix} lines.}~\reviewtwo{In Fig.~\ref{fig:qs_ch_intercorrelations_down}, we compare the velocities between {\siiv} and the hotter lines during the downflow phase in {\modch}. In general, the correlation between the velocities is very mild, with a few strong correlation cases ({\modch} {\fexv} upflow with {\siiv} upflow, {\oiv} and {\siiv} upflow for both the regions, {\fexv} downflow and {\siiv} upflow in {\modqs}, and {\fexv} \& {\oiv} downflow with {\siiv} downflow in {\modch}). The general lack of correlation is however evident in the downflow phase. }  
\end{enumerate}

~\review{We note that results (1), (2) and (3) are in consonance with established literature~\citep[for instance,][for a few references]{yokoyama_1996_jetsimulations,miyagoshi_2004_thermalconduction,daniel_2017_simulation}.}\review{The dynamics in {\modch}, for instance, are also reminiscent of simulations of emergence of twisted flux tube by~\cite{nobrega_2016_cooljet}, with the formation of a cool jet, a hot jet, and shocks.~\cite{nobrega_2016_cooljet} find the cool and hot jet forming on either side of the emerging flux tube. These differences between {\modch} and ~\cite{nobrega_2016_cooljet} may arise between the two setups due to different background topologies, resulting in reconnection onset in different environments. A qualitative comparison of the synthesized emission may be performed with \cite{daniel_2017_simulation}, who also perform synthesis in {\siiv} line. This work also finds jets travelling at $\approx100$ km/s of the same scale as the jets in our simulations. However, we also note the positions of the surge and hot jet on either side of the dome in the slanted background field case of \cite{daniel_2017_simulation}, which is opposite to what we ~\reviewtwo{find in our model}. }

Cool plasmoids in \ion{Ca}{2} H and K have been observed to move at $\approx35$ km/s~\citep{singh_2012_plasmoids,Rouppe2017plasmoids}. Cool, low velocity ``bright dots'', $\approx0.6\pm0.3$ Mm in size have also been observed \citep{peter_2019_plasmoids,tiwari_2022_plasmoids}. \cite{tiwari_2022_plasmoids} compare their observations with Bifrost simulations, and find that these dots form due to magnetic reconnection between emerging flux with pre-existing flux, very low in the atmosphere. Thus, these bright dots and plasmoids appear to satisfy some aspects of the plasmoids seen in our simulations. 

The flows in {\modch} are similar in magnitude to those observed in different spectral lines of spicular simulation by \cite{juan_2018_correlatedflows}. However, one crucial observation in our simulation is the presence of strong upflows in {\modch} near the boundaries of bright regions. \cite{juan_2018_correlatedflows} find the strong upflows to be co-spatiotemporal with intensity enhancements in lines like \ion{Fe}{12} and \ion{Fe}{14}.~\reviewthree{We do not find any such associations for upper transition region or coronal lines like {\neviii}, {\feix} or {\fexv}.} These non-co spatiotemporal intensity and velocity enhancements are more similar to some of the events observed by \cite{Conrad_2021_smallscaleupflows}.

Statistical analysis by considering the moments of the spectral lines as a function of {\bmag} was studied in {\ppsi} and {\ppmg} while correlations between velocities of different spectral lines were studied in {\ppmg}.~\review{We note that the results from the aforementioned papers are for CH and QS regions in general, and not any particular flux emergence scenario. However, the hypothesis presented in {\ppsi} and {\ppmg} attributes their results to interchange reconnection in CH and closed loop reconnection in QS, which are the primary jet generation mechanisms in our simulations.} Since we have performed a similar analysis with {\modqs} and {\modch}, we may perform a qualitative assessment of our results with these observations. The observations in {\ppsi} and {\ppmg} indicated a regular increase in line 
moments for chromospheric~\ion{Mg}{2},~\ion{C}{2} lines and the {\siiv} line with {\bmag}. {\siiv}, also synthesized in this work, shows similar dependence with {\bmag} in Fig.~\ref{fig:qs_ch_modb}. ~\review{ {\ppsi} and {\ppmg} however also find the CH intensities to be lower than QS intensities when these values are binned over {\bmag} -- an observation we do not find in these simulations. However we note that {\siiv} analysis may significantly depend on the realism of treatment of the chromosphere.}

{\ppsi} and {\ppmg} find an excess in upflows in CH over QS, with these upflows increasing with the formation temperature of the spectral line in consideration. We once again observe this signature in  Fig.~\ref{fig:qs_ch_modb} in {\siiv}. Furthermore, the dependence of line width on {\bmag} in {\modch} simulations (see Fig.~\ref{fig:qs_ch_modb}) are consistent with those observed by {\ppsi}.

{\ppmg} reports upflows in {\siiv} correlated with those in \ion{C}{2} and \ion{Mg}{2}, with these upflows in hotter lines larger in CH. While we do not synthesize the chromospheric lines, we consider similar correlations between the {\siiv} and other lines forming at larger temperatures.~\reviewthree{In {\modch},} we obtain these~\reviewthree{statistically significant} signatures in Fig.~\ref{fig:qs_ch_intercorrelations}(~\reviewtwo{first row}) -- not just for a couple of lines, but across lines spanning a range of temperatures~\review{confirming the coupling across the atmosphere}. However, correlated upflows between {\siiv} and \ion{Fe}{12} line inside a CH were observed by \cite{Conrad_2023_Smallscaleupflows}. These flows are at $\approx10$ km/s in both the lines, different from the flows seen in {\modch}. 

{\ppmg} also reports signatures of ``bidirectional flows'' between the downflows in chromospheric lines and upflows in the transition region lines. We find ~\review{mild correlations} between downflows in {\siiv} and upflows in other lines, and find strong upflows in CHs (again in Fig.~\ref{fig:qs_ch_intercorrelations}, ~\reviewtwo{last row}). It is rather interesting that the signatures observed approximately at the base of the transition region in {\ppmg}, and hypothesized to be seen throughout the atmosphere, are observed approximately in the transition region and the base of the corona in {\modch}.

{\ppmg} also find correlated downflows across all spectral lines. The chromospheric lines showed predominant downflows, even as a function of {\bmag}. However, we do not find any such signatures of correlated downflows in this work~\reviewthree{(see Fig.~\ref{fig:qs_ch_intercorrelations_down} third row)}, since our focus rests mainly on lines forming at temperatures higher than {\siiv}. %However, we note that the chromosphere is known to be predominantly downflowing~\citep[see, for example][]{Stucki_UVLinesSumer_Chs}~\reviewthree{, and since we do not simulate a re}

~\review{We can now evaluate the} hypothesis presented in  {\ppsi} and {\ppmg}~\review{with our simulations}. ~\review{{\ppmg} hypothesize that while closed loop reconnection only results in local filling of loops in QS, interchange reconnection in CHs enables correlated plasma motion along the atmosphere column. The upflows are accelerated in CHs with height, resulting in spectral signatures in hotter lines, while such a phenomenon does not occur in QS}.

~\review{In {\modch}, interchange reconnection results in the formation of a surge, wherein the cool plasma trapped in the rising loop is liberated. As the reconnection proceeds, the forward edge of the jet interacts with the ambient atmosphere, leading to a ``hot jet'' on top of the cool jet.~\review{Contrary to the hypothesis in {\ppmg}, the cool and hot upflows do not form due to a transformation of the cool jet into the hot jet. The two jets are correlated though, as the interaction of the forward edge of the cool jet with the ambient atmosphere results in the hot jet, and they both proceed in the same general direction. Line of sight integration results in an approximate co-spatial association between the two jets in {\modch}.} Hence, spectral synthesis suggests correlated upflows in {\modch} between the cool and hot jet (result (5)). In {\modqs}, reconnection occurs and results in the formation of the cool jet with the hot jet at the forward edge. However, the jet is not oriented in the direction of the observer. Hence, we do not see such correlated upflows in {\modqs}.} Furthermore, the bidirectional flows result from explosive reconnection that also sends a portion of the reconnected plasma towards the photosphere, a consequence of the `slingshot' effect.

~\review{The jet reconnection and dynamics occur high up in the atmosphere. However, we find the velocities and line widths to be correlated with {\bmag}, which is computed at $z=0$. Given the apparent non-local association between the photospheric field and spectral response of upper atmosphere, we would not have expected the correlations~\reviewtwo{in velocities and line widths of hot lines} as seen in \S.~\ref{sec:upflow} and ~\ref{sec:downflow}, especially Fig.~\ref{fig:qs_ch_modb} and ~\ref{fig:qs_ch_modb_down}. The existence of such a strong correlation can be explained by examining the magnetic field topology, seen for example Fig.~\ref{fig:ch_density_temp}. The emerging flux undergoes reconnection on the edge oriented approximately opposite to the background topology. This results in a surge along the edge oriented in approximately the same direction as the background field, reflecting into velocity--{\bmag} correlations. We also note that correlations do not imply causation. The {\bmag} at $z=0$ at the timestamp in consideration is not causally giving rise to the velocities, and they are related by magnetic topology.}

~\review{As the jets in {\modch} move outward, at some point the cool plasma starts falling back down. ~\reviewtwo{Along with the cool plasma, the hot plasma on top of it also moves back down.} These are seen as downflows in the hot lines in Fig.~\ref{fig:ch_velocity_xt}. However, the motion of the surge ``cleared out'' some of the overlying hot plasma, and as it falls back down~\reviewtwo{, it} results in a significant, transient, intensity reduction in the hot lines, seen in Fig.~\ref{fig:ch_netflux_xt}. Another supporting evidence comes from the co-spatio-temporal association between the reduced intensities and the strong downflows seen in Fig.~\ref{fig:ch_velocity_xt}. Here, we also find anti-correlated {\modch} intensities with {\bmag} in Fig.~\ref{fig:qs_ch_modb_down}, which is really a result of the strong velocity correlation seen in Fig.~\ref{fig:qs_ch_modb}. We further note that the strongly upflowing plasma simply falls back down along the field lines, and manifests as downflows strongly correlated with {\bmag} in the hot lines. While some aspects of these results are consistent with the observations in {\ppmg}, we note that {\ppmg} hypothesizes correlated downflows across all lines. We find that hot and cool plasma have essentially decoupled during the return phase. However, these results are not in tension with those in {\ppmg}, since those observations probe only the lower atmospheric flows. If we were to consider just the downflows in the hot lines from Fig.~\ref{fig:ch_velocity_xt}, we can clearly see (even without any binning) that the downflows are co-spatial and co-temporal across the different temperatures. Hence, the hypothesis extends only for a certain temperature regime, and not across the atmosphere.}

There are, however, a few caveats in performing the comparisons between our simulations with the observations in {\ppsi} and {\ppmg}. One of the most direct comparisons would be the velocities and line widths observed in {\siiv} versus those computed from this simulation. Our simulation underestimates both the velocities and the line width when compared to {\ppsi}. This suggests both additional mechanisms that drive the flows low in the atmosphere, and a more realistic chromosphere to capture various physical effects. This may need, for instance, ambipolar diffusion, which is known to the formation of faster spicules~\citep{juan_ambipolar}.~\review{We also note that a self-consistent, realistic granulation and a proper treatment of radiation~\citep[similar to, for instance][]{nobrega_2016_cooljet,daniel_2017_simulation,nobrega_2022_nullptreconn} would enable a more realistic simulation enabling quantifiable differences from observations.} We note also that the initial temperature profile must be more realistic, with realisitic values in the photosphere and chromosphere.~\reviewthree{We note that the higher temperature at z = 0 in this simulation leads to shallower stratification of the model atmosphere. A model with the photosphere at $\approx6000$ K would have a steeper stratification, resulting in (i). Lower densities in the atmosphere, and (ii). Stronger flux sheet field strength at equilibrium. We would expect these two factors to result in reconnection to occur at lower heights, since the onset criterion depends on $J/\rho$, which would be satisfied for at lower heights.} Furthermore, our simulation was 2.5D, without any treatment of radiative transfer in the chromosphere and lower atmosphere. We also note that {\ppsi} and {\ppmg} were probing much smaller spatial scales, which are not exactly studied in this simulation setup.%An accurate modeling of the dynamic phenomena would need a correct treatment of radiation, the inclusion of 3D scattering effects~\citep{leenarts_iris1}, and the production of a self-consistent corona with photospheric driving. Such 3D numerical models would appropriately let us understand the specific effect of emerging flux in CH and QS regions. Furthermore, the results in \cite{Upendran_C2, Upendran_2022_CHQS} need the reconnection to occur at much lower heights, which would need us to perform very small-scale flux emergence experiments for comparison.

Finally, the spectral response to the flux emergence dynamics has been performed for spectral lines that would be observed by MUSE and SOLAR-C (EUVST). The reconnection dynamics simulated in this work will leave strong imprints in the different spectral lines. The whole event lasts for $\approx400$ s, with a net spatial extent of $\approx20$ Mm. Each plasmoid is, of course, smaller ($\leq1$ Mm), and propagates over the time scale of 400 s. EUVST should be able to capture the dynamics of such events and enable a statistical comparison between CH and QS. On the other hand, MUSE would be crucial with observations in {\feix} and possibly {\fexv}, capturing the high-temperature dynamics. Thus, a concerted observation campaign, along with more realistic simulations building on this work would provide great insights into understanding solar wind emergence in CH and local heating in QS in a unified manner.

%------------------
\section{acknowledgments}
U.V. gratefully acknowledges support by NASA contracts NNG09FA40C (IRIS), and 80GSFC21C0011 (MUSE). U.V. would like to convey gratitude to Viggo Hansteen for suggesting the use of a hot plate in the top boundary to maintain coronal temperature. U.V. would also like to thank Anuj Mishra and Abhishek Rajhans for discussions on tackling numerical instabilities in code. U.V. would like to thank Mats Carlson, Philippe Bourdin, and Georgios Chintzoglou for discussions during the Hinode--16/IRIS--13 meeting at Niigata, Japan. U.V. would like to acknowledge ChatGPT v3.5 for automating the composition of the visualization code, that generates figures from simulation cubes and updates them on Weights and Biases, providing a remote access check on the various simulation runs. These figures were used for simulation monitoring purposes and for generating reports internally. \review{We would also like to thank the reviewer for their comments that improved the manuscript.} We would like to acknowledge the use of the CHIANTI data base. CHIANTI is a collaborative project involving George Mason University, the University of Michigan (USA), University of Cambridge (UK) and NASA Goddard Space Flight Center (USA). A part of this work appeared in the Ph.D dissertation of U.V~\citep{Vishal_PHD_thesis}

\software{Astropy~\citep{astropy}, Jupyter~\citep{jupyter},  Matplotlib~\citep{matplotlib},   Multiprocessing~\citep{multiprocessing}, Numpy~\citep{numpy_nature}, Scipy~\citep{scipy}, Sunpy~\citep{sunpy}, Xarray~\citep{hoyer2017xarray}, Weights and Biases~\citep{wandb}.}

%---------------------
\appendix
\section{Grid details}
\label{sec:app_1}
 The simulation grid is described in \S.~\ref{sec:sim_setup}. In the horizontal direction, we have a logarithmic grid at $x\leq40.3$ Mm and $x\geq80.6$ Mm. In a logarithmic grid, we first define $\Delta\xi^\pm$ as:
$$\Delta\xi^{\pm} = \pm\frac{1}{N}\log \left( \frac{ x_R+\vert x_L\vert -x_L}{\vert x_L\vert}\right),$$\reviewfour{where $\pm$ corresponds to increasing or decreasing mesh,~\reviewsix{$x_R$ and $x_L$ correspond to the rightmost and leftmost point of the patch in consideration, and N is the number of points in the \reviewfive{patch} }. The grid size at $i^{th}$ grid point is defined as:} $$\Delta x_i^{+} = \left(x_{i-1/2}+\vert x_L\vert - x_L \right)(10^{\Delta\xi^+} - 1), \Delta x_i^{-} = \left(x_{i-1/2}-\vert x_L\vert - x_R \right)(10^{\Delta\xi^-} - 1),$$ 
where $\Delta x_i^{\pm}$ correspond to the increasing and decreasing mesh respectively. \reviewfive{In our case, for example, the patch from ($x_L \sim80.6$ Mm, $x_R\sim121.51$ Mm) has an increasing grid with 15 cells, while the patch from ($x_L \sim0.3$ Mm, $x_R\sim40.3$ Mm) has decreasing grid with 15 cells. As we already note in \S.~\ref{sec:sim_setup}, we ~\reviewsix{have} a constant, regular grid spacing between $x\sim40.3$ Mm and $x\sim80.6$ Mm. We further note that the vertical grid is explained in detail in \S.~\ref{sec:sim_setup}.}

We depict the grid spacing at each coordinate value in Fig.~\ref{fig:sim_grid} for {\modch}. For the sake of consistency, we also include the grid in z direction in the right panel of Fig.~\ref{fig:sim_grid}. \reviewsix{We note that the vertical grid for {\modqs} differs from {\modch} in the location of base of transition region. This is marked by a change of the grid from constant to a stretched grid, as described in \S.~\ref{sec:sim_setup} and can be seen in Fig.~\ref{fig:sim_grid}.}
\begin{figure}
    \centering
    \includegraphics[width=\linewidth]{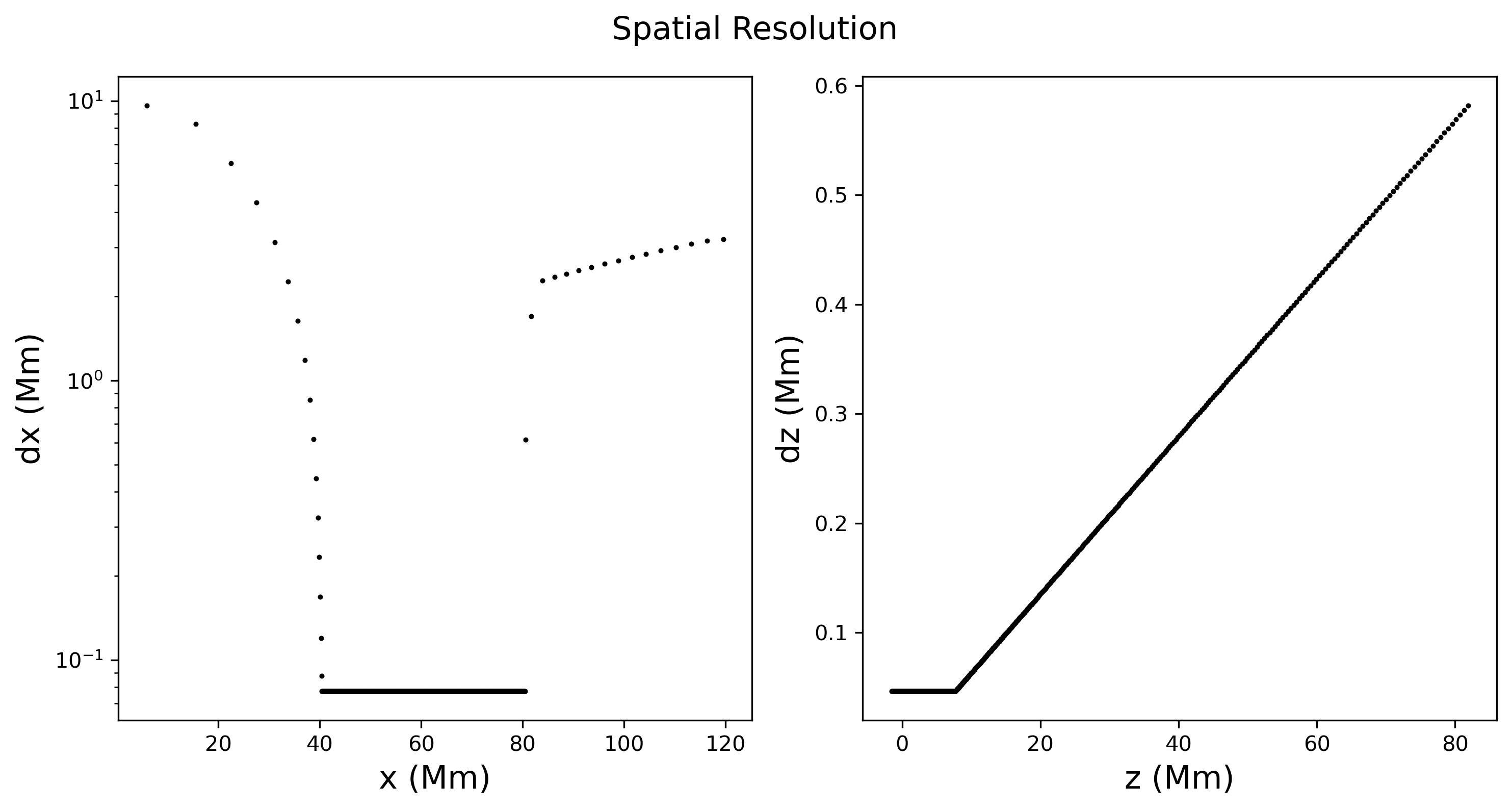}
    \caption{Simulation grid in x (left) and z (right) direction.}
    \label{fig:sim_grid}
\end{figure}

\bibliography{sample631}{}
\bibliographystyle{aasjournal}

\end{document}